\documentclass{article}

\usepackage[english]{babel}

\usepackage[letterpaper,
            top=1in,
            bottom=1in,
            left=1in,
            right=1in]{geometry}

\usepackage{fancyhdr}

\usepackage{enumitem}

\usepackage{tikz}
\usepackage{pgfplots}
\pgfplotsset{compat=1.10}
\usepgfplotslibrary{fillbetween}
\usetikzlibrary{patterns}

\usepackage{placeins}
\usepackage{color}
\usepackage{xcolor}
\usepackage{colortbl}
\definecolor{skyblue}{HTML}{93B7BE}
\definecolor{orange}{HTML}{E69F00}
\definecolor{DarkGrey}{HTML}{454545}
\definecolor{DarkGreen}{rgb}{0.1,0.5,0.1}
\definecolor{DarkRed}{rgb}{0.5,0.1,0.1}
\definecolor{DarkBlue}{HTML}{215CAF}
\definecolor{HighlightOrange}{rgb}{1, 0.647, 0}
\definecolor{HighlightRed}{rgb}{0.8, 0.2, 0.2}

\usepackage{natbib}
\setcitestyle{authoryear,round,citesep={;},aysep={,},yysep={;}}

\renewcommand{\cite}[1]{\citep{#1}}

\usepackage{amsmath}
\usepackage{amsfonts}
\usepackage{amsthm}
\usepackage{amssymb}
\usepackage{mathtools}
\usepackage{nicefrac} 
\usepackage{bbm}

\usepackage[utf8]{inputenc}
\usepackage[T1]{fontenc}
\usepackage{microtype}
\usepackage{csquotes}
\usepackage[scaled=0.90]{helvet}
\usepackage[scaled=0.85]{beramono}
\usepackage{textcomp}

\usepackage{graphicx}
\usepackage{subfigure}
\usepackage{booktabs} 
\usepackage{tabularray}
\usepackage{pdflscape}

\usepackage{hyperref}
\hypersetup{
	colorlinks=true,       
	linkcolor=DarkBlue,    
	citecolor=DarkBlue,    
	filecolor=DarkBlue,    
	urlcolor=DarkBlue,     
	pdftitle={},
	pdfauthor={},
}
\usepackage{url}            

\usepackage{algorithm}
\usepackage{algorithmic}

\theoremstyle{plain}
\newtheorem{theorem}{Theorem}[section]
\newtheorem{proposition}[theorem]{Proposition}
\newtheorem{conjecture}[theorem]{Conjecture}
\newtheorem{lemma}[theorem]{Lemma}
\newtheorem{corollary}[theorem]{Corollary}
\theoremstyle{definition}
\newtheorem{definition}[theorem]{Definition}
\newtheorem{assumption}[theorem]{Assumption}
\theoremstyle{remark}
\newtheorem{remark}[theorem]{Remark}

\newcommand{\cX}{\mathcal X}

\newcommand{\E}{\mathbb{E}}

\DeclareMathOperator{\Ind}{\mathbbm{1}}

\newcommand{\rank}{\textrm{rank}}

\usepackage{todonotes}

\usepackage{authblk}
\title{\Large Causal Inference from Competing Treatments}
\author[1]{Ana-Andreea Stoica}
\author[1,2]{Vivian Y. Nastl}
\author[1]{Moritz Hardt}
\affil[1]{Max Planck Institute for Intelligent Systems, T\"{u}bingen, Germany, and T\"{u}bingen AI Center, Germany}
\affil[2]{Max Planck ETH Center for Learning Systems}
\date{}

\begin{document}
    \thispagestyle{plain}
    \maketitle

    \begin{abstract}
        Many applications of RCTs involve the presence of multiple treatment administrators---from field experiments to online advertising---that compete for the subjects' attention. In the face of competition, estimating a causal effect becomes difficult, as the position at which a subject sees a treatment influences their response, and thus the treatment effect. In this paper, we build a game-theoretic model of agents who wish to estimate causal effects in the presence of competition, through a bidding system and a utility function that minimizes estimation error. Our main technical result establishes an approximation with a tractable objective that maximizes the sample value obtained through strategically allocating budget on subjects. This allows us to find an equilibrium in our model: we show that the tractable objective has a pure Nash equilibrium, and that any Nash equilibrium is an approximate equilibrium for our general objective that minimizes estimation error under broad conditions. Conceptually, our work successfully combines elements from causal inference and game theory to shed light on the equilibrium behavior of experimentation under competition.~\looseness=-1
    \end{abstract}

    \section{Introduction}
\label{sec1:introduction}

Randomized controlled trials (RCTs) have become ubiquitous in a variety of fields, to the point that they are ``not so much the `gold standard' as just a standard tool in the toolbox''~\cite{banerjee2016influence}. Application areas range from  A/B testing in industry to informing policy choices at a large scale~\cite{levy2007progress,alatas2012targeting,de2019rise}.~\looseness=-1

Given the abundance of experimentation, an agent who wishes to measure the effect of their treatment through an RCT may not be alone in doing so. Competing treatments arise when multiple agents share the population on which they wish to experiment. Subjects may then receive multiple treatments sequentially determined by some form of competition between the treatment administrators. Treatments that came before may modify the effectiveness of subsequent treatments. Precisely estimating effects therefore becomes challenging under competition: sequential treatments may pose the issue of external validity~\cite{duflo2006field} or mislead companies in their assessment of products (e.g. as the click-through-rate plummets for low-position content).~\looseness=-1

Our concrete running example is online advertising. Multiple advertisers wish to estimate the effectiveness of their campaign. However, due to competition between advertisers, their ad will be displayed at a certain position on screen. The lower the rank of the ad, the lesser its effectiveness~\cite{agarwal2011location}. Running the campaign, an advertiser collects data of impressions and clicks on ads displayed at different ranks. From this data, the advertiser wishes to answer the question:

\begin{quote} 
\centering 
\textit{What would have been the effect of the treatment had it been the first treatment applied?} 
\end{quote}

In other words, how effective is the ad if it is displayed in the top rank? Equipped with knowledge of this effect, an advertiser can distinguish between two scenarios: (1) an ad was not effective because it was mostly displayed at lower ranks, or (2) an ad was not effective because of its design. An advertiser can then take appropriate action to increase its effectiveness in subsequent campaigns. The advertiser faces two problems: one game-theoretic, the other statistical. First, how should the advertiser bid on subjects, anticipating that other advertisers will bid strategically, too? Second, how should data from different ranks be weighted optimally to obtain the best possible estimate of the causal effect?

Specifically, we show what the optimal estimation error is that an advertiser can achieve at equilibrium in a game that models the competition between advertisers. 
In other words, our work shows how to spend an experimental budget rationally on subjects when the goal is to estimate a causal effect optimally.~\looseness=-1

\subsection{Our contribution}

We introduce a novel model of causal inference under competing treatments. Our model combines elements from game theory and statistics to find an answer that neither toolkit can provide on its own. Specifically, how should treatment administrators allocate their budget so as to minimize statistical estimation error, while facing competition?~\looseness=-1

We assume that the causal effect of interest is a bounded quantity~$\tau$ that we interpret to be the effect of treatment had it come first. A number~$k$ of treatment administrators compete over a pool of~$n$ subjects. Administrators have budgets that they can spend by making bids on subjects. A probabilistic allocation rule determines the order in which each subject receives the $k$ treatments. We assume that the treatment effectiveness decays with the order in which the treatment is applied.
A treatment applied at position~$r$ has a causal effect $\tau_r<\tau.$ We assume that $\tau_r=\alpha_r \cdot\tau,$ where $\alpha_r$ is a known structural parameter of our model.

In our model, administrators bid strategically so as to achieve the smallest possible estimation error. The optimal estimation error depends on what we call the \emph{sample profile}, that is, the number of samples~$n_r$ at each rank~$r$. 
One of our key contributions is to approximate the optimal error objective by a more tractable objective that we call \emph{sample value}. The sample value corresponds to the weighted sum~\mbox{$S=\sum_r  n_r \cdot \alpha_r^2$.} The sample value quantifies the effective utility of the sample where each rank is discounted appropriately.

We prove that, in the typical regime where budgets are larger than~$k^2$, any Nash equilibrium in the sample value objective is also an approximate Nash equilibrium with respect to the optimal error objective. Using this approximation result, we can understand the competition over optimal error by instead analyzing the competition over sample value. This objective has many nice properties. 
We show that it has a pure Nash equilibrium, by finding a class of equilibria in closed form in which rational administrators spend their budget exhaustively in a way that maximally avoids competition.~\looseness-1

Conceptually, we find a fruitful bridge between causal inference and game theory.
On the one hand, causal inference literature tells us how to account for known position effects in an optimal way. On the other hand, game theory provides solution concepts for dealing with competition. A novelty is that we not only apply game theory to budget allocation strategies but rather to estimate causal effects. Our work opens the door for numerous interesting questions, which we discuss in Section~\ref{sec7:discussion}, alongside current limitations of our framework.~\looseness=-1

\subsection{Related work}
\label{sec2:relwork}

Several fields provide evidence that the position at which content is seen affects engagement, from early theories in experimental psychology~\cite{ebbinghaus1885position} to survey design~\cite{mcfarland1981effects}.
Because of position effects randomized experiments in development economics run the risk of reducing the external validity of sequential studies run on the same population~\cite{duflo2006field}.
Perhaps the most extensive work studies the digital space, where search results displayed lower on a webpage receive less engagement~\cite{ansari2003searchposition,teevan2008searchposition} and low-positioned ads are recalled less~\cite{varian2007position,agarwal2011location}.
In particular,~\citet{narayanan2015position} measure the effect of positioning ads under selection biases such as competition using regression-discontinuity, without modeling the competition for ranks explicitly.
Moreover,~\citet{hardt2022performative} define a framework for measuring the effect of position on market power acquisition.~\looseness=-1

Our work draws on this literature and models position effects that emerge from competition: agents who wish to run randomized controlled trials on a population encounter a position effect when competing with each other over `winning' the attention of subjects.
Such competition has been studied in game-theoretic settings.
In particular in applications like online advertising, pricing mechanisms have been developed in order to achieve stability in the market~\cite{aggarwal2006truthful,edelman2007internet}. A related line of work is that of position-based auctions.
Early works in this research line studied equilibria and truthfulness through designs that encode positions in the agents' valuations for individual users~\cite{varian2007position,athey2011position,borgers2013equilibrium}. Another recent line of work studies optimal platform design choices under proxy bidding (or `autobidding'), where agents' strategy space is limited to budget choices and targeting mechanisms~\cite{aggarwal2019autobidding,balseiro2021robust,conitzer2022pacing}.~\citet{ghosh2010expressive} study alternative designs for auctions with negative externalities generated by competing advertisers in online conversion-based auctions, without an explicit position-based model. Our work differs from this literature in the following ways: in our set-up, agents do not have an intrinsic value for particular individuals---a common auction-modeling choice in early position-based auctions---but rather a sample value that determines the estimation error of their. This brings our problem closer to the `autobidding' literature, yet, differs from it in that our optimization problem includes both bidding strategies and the choice of an optimal estimator for precise estimation of causal effects.~\looseness=-1

Our game-theoretic design is inspired by and directly generalizes the work of~\citet{maehara2015budget}. Their design optimizes the sample size of customers obtained by competing advertisers, in a line of work that bridges influence maximization with budget allocation~\cite{alon2012optimizing}. Whereas in~\citet{maehara2015budget}'s design there is no explicit rank (an advertiser either wins a customer or not, maximizing the size of their sample), our model generalizes theirs by including multiple positions in which advertisers may be shown to users. This generalized game may be of independent research interest, as we can characterize particular Nash equilibria.~\looseness=-1

Estimating treatment effects under equilibrium conditions has been studied in the case of unit interference~\cite{wager2021experimenting} and two-sided markets~\cite{johari2022experimental}. They consider dynamic models in which the treatment effect is dependent on the available market supply and unit's interactions, assuming a single treatment coming from a decision-maker (often a platform designer).

Adaptive experimentation is a rapidly growing area in estimating causal effects under sequential treatments.~\citet{hadad2021confidence} and~\citet{dwivedi2022counterfactual} provide statistical inference guarantees in the problem of estimating counterfactual means, without explicitly modeling multiple ranks or strategic interactions. Methods such as multi-armed bandits have been employed to provide guarantees on the efficacy of treatments through optimal data collection in the presence of position effects~\cite{lagree2016multiple,zhou2023bandit}. They primarily assume a \emph{single} learner. While bearing similarities in objectives (as our objective also directly optimizes for estimation error), our work differs from these by modeling data acquisition through competition that emerges from \emph{multiple} learners.~\looseness=-1
    \section{Games and objectives}
\label{sec3:setup}

We formally introduce a game that captures the setting of $k$ treatment administrators (also called admins) who compete over a set of~$n$ subject slots. Linking back to our online advertising example, the different admins are represented by different advertisers and the subject slots are positions at which users will see the displayed ads on an online platform. Conceptually, subject slots are placeholders for i.i.d.~samples drawn from a data-generating distribution. Administrators have varying budgets that they can spend exhaustively by bidding on the subject slots. The outcome of the game is an allocation of the subject slots to the $k$ administrators. Thus, each admin receives up to $n$ slots for displaying their ad. For each slot, they can run a treatment-control experiment and measure the effect of an ad. The formal definition of the data-generating process is given in Section~\ref{sec:causalinference}. We assume that the causal effect of the treatment at rank $1$ is given by a bounded quantity~$\tau$ with $|\tau|\le 1$.\footnote{The effect just needs to be bounded quantity. We assume it is bounded by $1$ without loss of generality.} We also assume that treatment at a higher rank $r>1$ is less effective so that~\looseness=-1

\[
\tau_r = \tau\cdot\alpha_r \quad\text{with}\quad 0<\alpha_r< 1\,.
\]
The discount factors~$\{\alpha_r\}$ are known structural parameters of our model. We discuss the case where the discount factors are not \textit{a priori} known in Section~\ref{sec:causalinference}. Samples at rank~$1$ are most valuable, but samples from lower ranks can still be useful. A treatment administrator bids strategically so as to minimize their mean squared error in estimating the causal effect~$\tau.$ To fully specify the game, we need the following ingredients:~\looseness=-1
\begin{itemize}
    \item We can represent the bids as a nonnegative integer matrix $\mathbf{x}\in\mathbb{N}_{\geq 0}^{n\times k},$ where $\mathbf{x}_{ai}$ is the bid of admin~$a$ on subject slot~$i.$ We denote the budget of admin~$a$ by $B^{(a)}=\sum_i \mathbf{x}_{ai}.$ An admin's strategy is a potential bid allocation over user slots. We denote the set of all possible strategies of an admin $a$ by $\mathcal{D}_a$.
    \item An allocation rule $\mathcal{A}$ maps bids to an allocation of subject slots. That is, given a set of bids, the allocation rule determines an assignment of ranks for each admin. A rank assignment of a particular admin is a tuple $\overline{r} = (r_1, r_2, \cdots, r_n)$, where $r_i$ is the rank assigned for subject slot $i$. Under a probabilistic rule, we draw the rank assignment from the distribution given by the randomized allocation. The \emph{sample profile} is just the count of the number of slots at each rank: $\overline{n} = (n_1, n_2, \cdots, n_{k})$, where $n_r = \sum_i \Ind \{r_i = r\}$. We denote the total amount of slots obtained by admin $a$ by $n^{(a)}=\sum_k n_k$. The specific allocation rule we define later is randomized.
    \item The utility function~$f_a\colon\mathcal{D}\to\mathbb{R}$ of admin~$a$ maps a tuple of strategies from~$\mathcal{D} = \mathcal{D}_1 \times \mathcal{D}_2 \cdots \mathcal{D}_k$ to the utility of admin~$a$ given the strategies of all admins. 
\end{itemize}
Our goal is to understand the Nash equilibria of the game. These are the strategies in which each admin is unilaterally maximizing her utility. In other words, admins are simultaneously best responding to each other. Specifically, we want to understand what estimation error admins can hope to achieve at equilibrium.~\looseness=-1 

\subsection{Objective design}
\label{sec:objdesign}
Given an assignment of ranks $\overline{r}$ drawn from the distribution induced by the allocation function $\mathcal{A}$, an administrator has to decide how to use the samples so as to optimally estimate the causal effect~$\tau$. This corresponds to solving the optimization problem, from the perspective of a single admin:~\looseness=-1
\begin{equation}
    \inf\limits_{\widehat{\tau}} \E_{Z\sim\mathcal{X}(\overline{r})} \left[ \left( \widehat{\tau}(Z) - \tau \right)^2\right]
    \label{eq:errorminimizationobjective}
\end{equation}
over all possible estimators~$\hat\tau.$ Here, $\mathcal{X}(\overline{r})$ denotes the data-generating distribution under the realized rank assignment~$\overline{r}$, defined as: given a population $\mathcal{X}$ and a ranking assignment $\overline{r}$, a subject is drawn independently at each subject slot $i$ from the interventional distribution $\mathcal{X}^{\textrm{do}(\textrm{rank} = r_i)}$, where $r_i$ is the rank that the admin obtained at subject slot $i$ (coordinate $i$ in the assignment rank vector $\overline{r}$). The data $Z$ drawn from $\mathcal{X}(\overline{r})$ is the collection $Z_i = \left(r_i, T_i, Y_i\right), i\in[n^{(a)}]$, where $T_i$ is the treatment assignment and $Y_i$ denotes the outcome of a subject sampled for slot $i$ at rank $r_i$. We formalize the data-generating process, together with the modeling choices for the outcome and treatment assignment in Section~\ref{sec:causalinference}. 
Recall that treatment effects vary with the rank at which an admin's campaign is run, and therefore the data-generating distribution depends on the rank assignment. 

\paragraph{Estimation error objective.} Solving objective~\eqref{eq:errorminimizationobjective} directly is a difficult task.
We therefore replace the optimal error with a minimax lower bound on the error.
We prove (in Section~\ref{sec:causalinference}) that for any estimator $\widehat{\tau}$, there exists a model instance $M$ of the data-generating distribution $\mathcal{X}(\overline{r})$ such that the estimation error is lower bounded by
    \begin{equation}
        \min \left( c \cdot\left( \textstyle\sum_{r = 1}^{k} n_r \cdot \alpha_r^2 \right)^{-1},1 \right)
    \label{eq:minimax-lowerbound-sample-profile}
\end{equation}
for a constant $c>0$. In our subsequent analysis, we set w.l.o.g. $c=1$.\footnote{The value of the constant $c$ does not change the results: as all admins minimize the error by selecting the best sample distribution, the constant $c$ can be ignored in this objective. We obtain identical results using the expression $\min \left(\left( \textstyle\sum_{r = 1}^{k} n_r \cdot \alpha_r^2 \right)^{-1},1/c \right)$ in our objectives. Our results only rely on this expression being bounded by a constant.}
When the number of samples obtained by an admin approaches $0$, the first term in the bound expressed in equation~\eqref{eq:minimax-lowerbound-sample-profile} tends to infinity. However, in this case, since the causal effect is bounded, $|\tau|\le 1,$ the admin's mean squared error is also bounded by $1$ (this bound is given by choosing the constant estimator, for example). We provide a detailed discussion of this property in the Appendix.
We show (in Section~\ref{sec:causalinference}) that the minimax lower bound in \eqref{eq:minimax-lowerbound-sample-profile} is attainable up to constant factors. For example, we can use an optimally weighted average of Horvitz-Thompson treatment effect estimators at each rank, which render an estimation error of $\mathcal{O} \left( \left(\sum_{r = 1}^k n_r \cdot \alpha_r^2 \right)^{-1}\right)$. 
This motivates defining the \textbf{estimation error objective} as
\begin{equation}
    f_a(\mathbf{x})=-\E_{\overline{r} \sim \mathcal{A}(\textbf{x})}\min \left(  \E_{\mathcal{X}(\overline{r})}\left(\textstyle\sum_{r = 1}^k n_r \cdot \alpha_r^2\right)^{-1}, 1 \right).
\label{eq:errorminimizationobjective_sampleprofile}
\end{equation}
The negative sign serves the purpose of turning an estimation error minimization objective into a maximization objective. In essence, the estimation error objective captures an admin's goal of minimizing their estimation error (up to constant factors) in expectation over the sample profile distribution given by the allocation rule.~\looseness=-1

\paragraph{Sample value objective.} Even the estimation error objective is not necessarily tractable. In particular, it appears difficult to reason about Nash equilibria with respect to the estimation error objective. One of our main contributions is to relate the estimation error objective to a tractable objective. Given a sample profile, we call the weighted sum $\sum_r n_r \cdot \alpha_r^2$ the \emph{sample value} given by a rank allocation $\overline{r}$. The \textbf{sample value objective} is defined as
\begin{equation}
    f_a(\mathbf{x})=\E_{\overline{r} \sim \mathcal{A}(\textbf{x})} \E_{\mathcal{X}(\overline{r})} \left[ \textstyle\sum_{r = 1}^k n_r \cdot \alpha_r^2\right].
    \label{eq:samplevalueobjective}
\end{equation}
Intuitively, the sample value objective and the estimation error objective are linked, as they are both a function of the  sample value $\sum_{r= 1}^k n_r \cdot \alpha_r^2$, obtained by an admin through a strategy choice. Formally, we can show that for typical parameter settings, any Nash equilibrium in the sample value objective is an approximate Nash equilibrium in the estimation error objective. Moreover, we can characterize a set of Nash equilibria in the sample value objective.

\subsection{Main results}
\label{sec:mainresults}
\begin{figure*}
    \centering
    \includegraphics[width=0.6\textwidth]{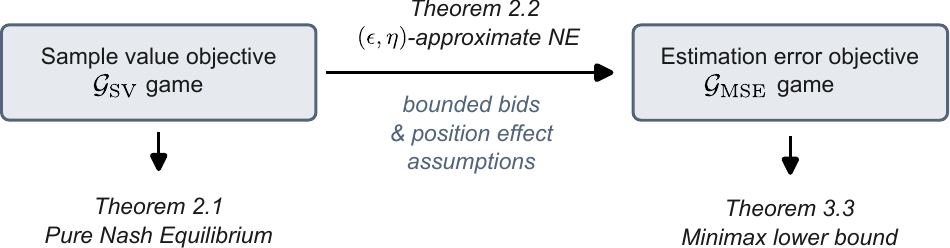}
    \caption{Illustration of the objective set-up and results.}
    \label{fig:objectives_illustr}
\end{figure*}

We denote by $\mathcal{G}_{\textrm{MSE}}$ and $\mathcal{G}_{\textrm{SV}}$ the games associated to the estimation error objective and the sample value objective, respectively. Figure~\ref{fig:objectives_illustr} illustrates our setup and results: through a causal inference analysis, we find an explicit expression for the minimax lower bound of an estimator as a function of the sample value (Section~\ref{sec:causalinference}); the sample value objective's associated game can be solved through finding a Nash equilibrium (Theorem~\ref{prop:negame}), described in detail in Section~\ref{sec:gametheory}, and any such solution approximates the error minimization objective under mild assumptions (Theorem~\ref{thm:approxthm}), which is the main result of our paper.~\looseness=-1

\begin{theorem}    
    A budget allocation game $\mathcal{G}_{\mathrm{SV}}$ with multiple treatment administrators has a pure Nash equilibrium for all sequences $(\alpha_r)_r$ such that $\alpha_r \in (0,1)$ and $\alpha_1 = 1$. Moreover, for $B^{(a)} \leq n$, an allocation in which all treatment administrators split their entire budget uniformly on subjects and maximally avoid competition on subjects is always a Nash equilibrium.
    \label{prop:negame}
\end{theorem}

\begin{theorem}
    Any Nash equilibrium for the $\mathcal{G}_{\mathrm{SV}}$ game with $k$ treatment administrators and $n$ subjects is an $(\epsilon, \eta)$-approximate Nash equilibrium for the $\mathcal{G}_{\mathrm{MSE}}$ game with
    \[
    \epsilon = O\left(\frac{k}{\sqrt{B}}\right)
    \quad\text{and}\quad
    \eta = O\left(\exp(-B/k^2)\right)\,,
    \]
    where $B=\min_a B^{(a)}$ and the maximum bid per subject slot is bounded.
\label{thm:approxthm}
\end{theorem}

In particular, if all administrators have budget $B^{(a)}=\omega(k^2)$, we get a $(o(1),o(1))$-approximate Nash equilibrium.
The proofs for all results are detailed in the Appendix.

\paragraph{Nash equilibrium.} We assume that players act rationally and selfish in maximizing their utilities.
We assume this is a complete information game (i.e. players know each other's utility functions). This set-up is common in game designs applicable in online advertising settings, as motivated by prior works~\cite{maehara2015budget,varian2007position}. We employ the well-known notions of \emph{pure} and \emph{approximate} Nash equilibrium. In short, a strategy set $\textbf{x}$ is a pure Nash equilibrium~\cite{nash1950equilibrium} if no deviation from it can increase utility for any admin (i.e. every admin's strategy is a best response). The concept of $(\epsilon,\eta)$-approximate equilibria has been widely used in finding approximate solutions~\cite{alon2012optimizing,roughgarden2016twenty}. In its essence, it assumes that an admin will not change his strategy if his current one is an approximate best response.~\looseness=-1

\begin{definition}
    A strategy set $\textbf{x} \in \mathcal{D}$ is an $(\epsilon, \eta)$-approximate Nash equilibrium for a utility-maximization game $\mathcal{G}$ with non-positive utility function $f_a: \mathcal{D} \rightarrow \mathbb{R}_{-}$ if for all players $a$, 
    \begin{equation}
        f_a(\textbf{x}_a, \textbf{x}_{-a}) \geq (1 + \epsilon) \cdot f_a(\textbf{x}_a', \textbf{x}_{-a}) - \eta, \forall \textbf{x}_a' \in \mathcal{D}_a
    \end{equation}
    where $\mathcal{D}_a$ denotes the strategy set of player $a$ and $\epsilon, \eta > 0$. An equivalent definition for a utility minimization game with non-negative utilities occurs for $\eta < 0$.~\looseness=-1
\label{def:approxobj}
\end{definition}

\paragraph{Allocation rule:} We use a probabilistic allocation rule that generalizes a model proposed by~\citet{maehara2015budget}. We defer a formal definition to Section~\ref{sec:gametheory}. Intuitively, each administrator has a chance to win the first rank of any given subject slot with a probability that is an increasing function of the administrator's bid for the slot. After determining a winner for rank~$r,$ the allocation rule proceeds to the next rank until all ranks have been allocated. The main properties of the allocation rule are that the probability for an admin of winning a rank is component-wise concave and increasing in the bid, which facilitates characterizing a Nash equilibrium. Furthermore, the allocation rule benefits from the property that the sample value does not grow in expectation slower than $B^{(a)}/k$: $\mathbb{E}\left[ \sum\limits_{r = 1}^{k} n_r \cdot \alpha_r^2 \right] = \Omega\left( B^{(a)} / k\right)$ under a Nash equilibrium for $\mathcal{G}_{\textrm{SV}}$. This allows the sample value variable to concentrate around its mean in the regime where budgets are larger than $k^2$.~\looseness=-1

\vspace{0.2in}

\noindent \emph{Proof sketch for Theorem~\ref{thm:approxthm}: } The intuition behind the proof relies on a few properties of the model and the allocation rule. Mainly, the allocation allows the sample objective to concentrate around its expectation with high probability in the bounded parameter regime. This allows us to show that if an allocation is indeed an optimal solution for the sample value objective, it will be approximately optimal for the estimation error objective. The `failure' to concentrate probability determines a distance $\eta$ from the utility of any other allocation; through a market-efficiency interpretation, it determines the necessary overhead above which an admin would be incentivized to change its strategy.
We discuss alternative allocation rules and connections to common auction designs in Section~\ref{sec7:discussion}, noting that our analysis is not contingent on the specific choice of an allocation rule, but rather on the properties mentioned in the proof sketch.

    \section{A causal inference analysis of the error minimization objective}
\label{sec:causalinference}

We treat the estimation problem from the perspective of a particular admin $a$. For simplicity, we drop in the notation the index $a$ when there is no confusion.
We define a population distribution $\mathcal{X}$ from which subjects will be independently drawn and used in the estimation problem for the admin. The admin runs an RCT on each subject and uses the data they gathered to estimate a causal effect. We employ classic modeling choices from the causal inference literature to model the outcomes and treatment assignments of subjects under an RCT.~\looseness=-1

\paragraph{Data generating process.} We detail the data-generating process for an admin $a$, briefly described in Section~\ref{sec3:setup}. The process consists of two parts, sketched in Figure~\ref{fig:causalinference} in the Appendix:~\looseness=-1 

\begin{enumerate}
    \item Admin $a$ competes over the $n$ subject \textit{slots} with the other admins by placing bids. A set of bids $\textbf{x}$ placed by the admins induce a distribution of rank allocations $\mathcal{A}(\textbf{x})$, where $\mathcal{A}$ is the rank allocation function. We sample a rank assignment $\overline{r}$ from $\mathcal{A}(\textbf{x})$ for admin $a$. The rank assignment $\overline{r}$ is a vector containing the assigned rank of admin $a$ for each subject slot.
    The sample profile $(n_1, n_2, \cdots, n_{k})$ gives the number of subjects drawn at each rank, on whom the admin will run their campaign. In total, the admin obtained $n^{(a)}=\sum_k n_k$ slots out of the total $n$ slots.\footnote{Each admin may obtain a different number of data slots. For example, if an admin bids $0$ on a slot, they will not obtain any rank for that slot; an admin bidding all $0$s will not obtain any slots, and therefore $n^{(a)} = 0$ for him.}~\looseness=-1

    \item The subjects are drawn independently at each slot from the interventional distribution $\mathcal{X}^{\textrm{do}(\textrm{rank} = r_i)}$, where $r_i$ is the rank that admin $a$ obtained at subject slot $i$ and treatment or control are allocated in a randomized manner (through an RCT). The data the admin collects consists of the tuples $(r_i, T_i, Y_i)$: rank $r_i$, treatment assignment $T_i$, and outcome $Y_i$ of each subject $i\in[n^{(a)}]$. As the RCT eliminates the influence from any possibly confounding factors, the treatment effect at a specific rank $r$ is identifiable~\cite{pearl2009causality}. 
\end{enumerate}

We point out that the budget allocation happens \textit{prior} to the sampling process. In particular, the individuals for each slot are not yet realized when the ranking assignment occurs. This process ensures that the individuals are independently drawn. We use potential outcomes to model the outcome of each individual sampled after the second step of the data-generating process occurs, described as follows.

\paragraph{Potential outcomes model.}
We model the outcome of a subject $i\in n^{(a)}$ sampled from $\mathcal{X}^{\textrm{do(rank} = r_i)}$ as 
\begin{equation}
    Y_i = c_{i,0} + c_{i,1}^{(r_i)} \cdot T_i + \varepsilon_i, \quad r \in [k]
    \label{eq:pot-outcome-one-rank}
\end{equation}
where $T_i$ denotes the treatment assignment of subject $i$ by the admin. We interpret $T_i = 1$ as assigning subject $i$ to the treatment group, and $T_i = 0$ to the control group. We model $T_i \sim \textrm{Ber}(q_i)$, for $q_i \in [q, 1-q]$ for a constant $q \in (0,0.5], \forall i \in [n^{(a)}]$. Essentially, we assume the admin implements a non-uniform Bernoulli randomized design.
The observational noise $\varepsilon_i$ is drawn from a normal distribution and does not depend on the subject's rank,
\begin{equation*}
    \varepsilon_i \sim N(0,\sigma^2).
\end{equation*}
We denote the potential outcome of subject $i$ for being in the treatment and control groups by $Y_i(1)$ and $Y_i(0)$, respectively.~\looseness=-1

\paragraph{Estimand: ATE but only for the first rank.} We are interested in the treatment effect an administrator would have obtained if they had experienced no competition from other administrators. 
In reality, competition causes an admin to show up at lower positions for some slots. The lower the rank for a slot, the smaller the treatment effect. 
The causal effect of interest is thus the \textit{average treatment effect had it been the first treatment applied}.
We formally define the estimand as
\begin{equation}
    \begin{gathered}
        \tau = \mathbb{E}[Y_i(1) - Y_i(0) | \textrm{do}(\rank = 1)],
        \label{eq:estimand_rankr}
    \end{gathered}
\end{equation}
where the expectation is taken over the intervened population $\mathcal{X}^{\textrm{do}(\textrm{rank} = 1)}$ that always encounter the content of admin $a$ first.
\begin{assumption}[Bounded effect]
    We assume that the treatment effect at the first rank $\tau$ is bounded \mbox{with
    $
        |\tau| \leq 1.
    $}
    \label{assumption:effectboundrank}
\end{assumption}

Given that a treatment administrator may acquire data from subjects at different ranks, we also define the average treatment effect of an administrator at rank $r\in [k]$ as
\begin{equation} 
    \tau_r = \mathbb{E}[Y_i(1) - Y_i(0) |  \textrm{do}(\rank = r)].
\end{equation}
Note that we identify the treatment effect at rank $r$ by \mbox{$\tau_r = \frac{1}{n_r} \sum\limits_{i:r_i=r} c_{i,1}^{(r)}$} if $n_r\geq 1$.

\begin{assumption}[Position effect]
    We formalize the assumption that the treatment effect at a lower rank is lower than at the first rank as 
    \begin{equation}
    \begin{gathered}
        \tau_r = \alpha_r \cdot \tau, \forall r \in [k]
    \end{gathered}
    \end{equation}
    with ${(\alpha_r)}_{r}$ satisfying $\alpha_1=1$ and $\alpha_r \in (0,1)$ for all $r \in [k]$.
    \label{assumption-scaled}
\end{assumption}

An admin wishing to estimate $\tau$ may of course just use the data from samples at rank $1$ ($n_1$ samples). However, the position effect assumption allows the admin to identify the treatment effect at the first rank \textit{from} the treatment effect at rank $r\in [k]$ and thus, to use the entire sample profile, 
\begin{equation*}
\begin{gathered}
    \tau = \frac{\tau_r}{\alpha_r}.
\end{gathered}
\end{equation*}
We refer to
\begin{equation}
    \tau^{(r)} := \frac{\tau_r}{\alpha_r}
\end{equation}
as the $r$-estimand.

\begin{remark}
Estimating the treatment effect at rank $1$ is a powerful primitive, as it allows one to extrapolate the effect at any \emph{distribution} over ranks, given the discount factors $(\alpha_r)_r$.
As mentioned in the introduction, this knowledge is important in order for an admin to gain insight into the quality of their treatment: was a lower effect observed because of a lower rank, or because of the treatment itself? Equipped with this knowledge, an admin can choose the best course of action: increase their bid or improve on the treatment design (e.g. in the case of online advertising, change the design of an ad). We emphasize that the admin has prior knowledge of the parameters $(\alpha_r)_r$ in this analysis. We discuss the case of jointly estimating $\tau$ and $(\alpha_r)_r$ later in this section.
\end{remark}

\subsection{An optimality argument}
The problem an admin is solving is now that of optimally using their collected data in order to achieve a minimum estimation error. Exactly \textit{how} an admin should combine their samples from different ranks towards minimizing the estimation error is non-trivial.
We show however that the estimation error of any possible estimator is bounded below by a function of the sample value $\sum_{r=1}^k n_r \cdot \alpha_r^2$.

To formalize this result, we denote the set of our potential outcomes model instances as $\mathcal{M}$. In particular,
\begin{equation}
    \mathcal{M} = \left\{\left(c_{i,0},c_{i,1}^{(r_i)}\right)_{i\in[n^{(a}]} : |\tau| \leq 1,  \tau_r = \alpha_r \cdot \tau \right\}.
\end{equation}

\begin{theorem}[Minimax Lower Bound]
    For any estimator $\widehat{\tau}$, there exists an instance $M \in \mathcal{M}$ such that the minimax squared error is bounded below by 
    \begin{equation*}
        \mathbb{E}_M[(\widehat{\tau} - \tau)^2] \geq \min \left(\frac{\sigma^2}{16 \cdot (1-q) \cdot \sum_{r=1}^k n_r \cdot \alpha_r^2},1\right).
        \label{eq:minimax-lowerbound}
    \end{equation*}\label{thm:minimax-lowerbound}
\end{theorem}

We provide the detailed proof in the Appendix. The proof relies on Le Cam's method~\cite{lecam1973minimax} for hypothesis testing and on leveraging the assumption of independence between individuals. The particular choice of an optimal estimator for $\tau$ is beyond the scope of this paper.
Instead, we showcase an estimator that achieves the minimax lower bound (up to a constant).

\paragraph{Estimator examples.} We find examples of estimators that aggregate the effect at each rank and obtain an error upper bound of $\mathcal{O} \left( \left(\sum_{r = 1}^k n_r \cdot \alpha_r^2 \right)^{-1}\right)$. Estimators that aggregate the effect at each rank are often used in meta-analysis studies or stratified experiments~\cite{hartung2011statistical, de2022trading}.
We refer to these as \emph{decomposable} estimators. Formally, a decomposable estimator can be written as $\widehat{\tau} = \sum_{r = 1}^{k} \omega_r \cdot \widehat{\tau}_r$ with $\widehat{\tau}_1^{(r)}$ an estimator for the $r$-estimand for $r\in[k]$. For unbiased estimators $\widehat{\tau}_1^{(r)}$, the well-known inverse variance estimator~\cite{Markowitz:1952,Markowitz:1959,cochran1954combination,shahar2017minimizing} provides a closed form solution for optimally weighing the $r$-estimators while preserving unbiasedness. The optimal weights $\omega^{*}$ are obtained by solving a variance minimization problem through Lagrange multipliers,
 \begin{equation}
     \omega_r^{*} = \frac{1}{\textrm{Var}\left(\widehat{\tau}_1^{(r)}\right)} \cdot \frac{1}{\textstyle\sum_{r = 1}^{k} 1 / \textrm{Var}\left( \widehat{\tau}_1^{(r)}\right)}.
 \end{equation}
 An immediate corollary is that the minimum MSE achieved for weights $\omega^{*}$ is 
    \begin{equation}
        \mathbb{E}\left[ \left( \tau_1 - \widehat{\tau}_{\omega^{*}}\right)^2\right] = \frac{1}{\sum\limits_{r = 1}^{k} 1 / \textrm{Var}\left( \widehat{\tau}_1^{(r)}\right)}.
        \label{eq:optimalmseweights}
    \end{equation}
Within our potential outcomes model, there are unbiased estimators for the $r$-estimand that achieve an error $\mathbb{E}\left[\left(\widehat{\tau}_r - \tau_r \right)^2\right] = \mathcal{O}\left( n_r^{-1}\right)$. Any such estimators, weighted optimally, would attain a total error bounded by $\mathcal{O} \left( \left(\sum_{r = 1}^k n_r \cdot \alpha_r^2 \right)^{-1}\right)$, as we will further argue. Given the parametric assumption on the sequence $(\alpha_r)_r$, we know that $\alpha_r$ amplifies the variance of the $r$-estimator by a squared term,~\looseness=-1
\begin{equation}
    \textrm{Var}\left(\widehat{\tau}_1^{(r)}\right) = \frac{\textrm{Var}\left(\widehat{\tau}_r\right)}{\alpha_r^2}  
\end{equation}
and thus, $\mathbb{E}\left[\left( \widehat{\tau}_1^{(r)} - \tau_1^{(r)}\right)^2 \right] = \mathcal{O} \left( (n_r \cdot \alpha_r^2)^{-1}\right)$. It then immediately follows from equation~\eqref{eq:optimalmseweights} that the estimation error of the inverse variance estimator $\widehat{\tau}_{\omega^{*}}$ is
\begin{equation}
    \mathbb{E}\left[ \left( \tau_1 - \widehat{\tau}_{\omega^{*}}\right)^2\right] = \mathcal{O} \left( \left( \sum\limits_{r = 1}^{k} n_r \cdot \alpha_r^2 \right)^{-1} \right)
\end{equation}

As examples of unbiased estimators for the $r$-estimand, the OLS estimator and the Horvitz-Thompson estimator~\cite{horvitz1952generalization} have variance of order $\mathcal{O}(n_r^{-1})$. The OLS estimator attains the lowest actual MSE, including constants, for unbiased linear estimators (see Gauss-Markov Theorem in~\citet{johnson2002applied}), while the Horvitz-Thompson estimator
is preferred in more general models that include unequal treatment probability (unequal $q_i$ across subjects). We do not reproduce the bound computation for the OLS estimator as it is a known result, but we detail, for the interested reader, properties of the Horvitz-Thompson estimator in our potential outcomes model.~\looseness=-1

We define the Horvitz-Thompson estimator formally for estimating the effect at rank $r$,
\begin{equation}
    \begin{gathered}
        \widehat{\tau}_r = \frac{1}{n_r} \cdot \sum\limits_{i:r_i = r} \left(\frac{Y_i \mathbbm{1}\{T_i = 1 \}}{\mathbb{P}(T_i = 1)} 
        - \frac{Y_i \mathbbm{1}\{T_i = 0\}}{\mathbb{P}(T_i = 0)}\right)
    \end{gathered}
    \label{eq:horvitzthompson_estimator_rankr}
\end{equation}
Similar to the OLS estimator, it is unbiased and has a variance of order $\mathcal{O}(n_r^{-1})$ (albeit with a larger constant factor). We show these properties in Propositions~\ref{prop:taurisunbiased} and~\ref{prop-hterror-ranked}.
\begin{proposition}
    The estimator $\widehat{\tau}_r$ is an unbiased estimator for the average treatment effect $\tau_r$ at rank $r$.\label{prop:taurisunbiased}
\end{proposition}
\begin{proposition}
    The error of the rank $r$ estimator $\widehat{\tau}_r$, $\mathbb{E}\left[\left(\tau_{r} - \widehat{\tau}_{r}\right)^2 \right]$, can be upper bounded by
    \begin{equation}
    \begin{gathered}
        \frac{1}{n_r} \cdot \left[\frac{\sigma^2}{q(1-q)} +  Y_{\mathrm{max}}^2 \cdot \left( \frac{(1-q)^2 + q^2}{q(1-q)} + 2 \right) \right]
    \end{gathered}
    \label{eq:mse-ranked-upperbound}
\end{equation}
    with $Y_{\mathrm{max}} := \max\limits_{i \in [n^{(a)}]} \left|c_{i,0}+c_{i,1}^{(r_i)}\right|$.\label{prop-hterror-ranked}
\end{proposition}

These results show that we can find an optimal estimator up to constants. While we do not require decomposable estimators in order to obtain the minimax lower bound on the error in Theorem~\ref{thm:minimax-lowerbound}, they provide a good intuition for the form of the sample value expression. The intuition is that whereas in a classical non-competitive causal inference task, the sample value is just a function of the sample size, the $\alpha_r$ downgrade in treatment effect from rank $1$ to rank $r$ scales down the sample value of rank $r$ by $\alpha_r^2$. Thus, we interpret the sum $\sum_{r = 1}^{k} n_r \cdot \alpha_r^2$ as the sample value that an admin has in order to estimate a treatment effect.
The choice of different strategies offers an admin the possibility of \textit{optimizing} the sample value.

\subsection{Estimating the discount factors}
\label{sec:estimatingalpha}

In our analysis so far, we assumed that the discount factors $(\alpha_r)_r$ are known by the admins. The parameters are perhaps given by a central platform or estimated in previous studies.\footnote{For example, Google reports the decay rate in click-through-rates with lower positions in online advertising~\cite{googlectrrates}.} In settings where the discount factors $(\alpha_r)_r$ are not \textit{a priori} known, they may be estimated from historical data. Related set-ups have utilized the EM algorithm in estimating the position bias in online advertisement problems~\cite{Chuklin:2016cm,dempster1977maximum}. In many applications of multiple treatments, such as online advertisement, the estimated discount factors do not depend on the particular admin but rather on the display position~\cite{lagree2016multiple}. In these cases, a pilot experiment is usually run by randomizing results or performing pairwise comparisons~\cite{joachims2017unbiased,wang2018position}, in order to get good estimates of the discount factors. The estimated discount factors are then used in subsequent campaigns, under the assumption that they do not need to be re-estimated.~\looseness=-1

For completeness, we show that estimating the discount factors jointly with the treatment effect through data splitting does not reduce the estimation error, even in the most simple case: take a pilot experiment in which an online platform has collected data on an admin's treatments at different ranks, with $n_r$ samples at each rank. Without prior knowledge of the position effect parameters, the admin has two choices in estimating $\tau$: (1) use just the data where their treatment was shown on the first rank, or (2) split the data at all positions, using part of it for estimating the position effect parameters $(\alpha_r)_r$, and then using these estimates together with the rest of the data for estimating the treatment effect $\tau$. Even if the admin has to split the data available in (2), he might still have more samples for the treatment estimation task than in scenario (1). We show in a simple argument that the treatment effect estimation error is not reduced in (2) as compared to (1), despite the fact that the admin is using more data in (2) and no matter \emph{how} the split is being done (Lemma~\ref{prop:jointestimation} in the Appendix).~\looseness=-1
    \section{Nash equilibria for the sample value game}
\label{sec:gametheory}

We define a probabilistic allocation rule $\mathcal{A}_{\textrm{prob}}$ and prove Theorem~\ref{prop:negame} by characterizing a set of pure Nash equilibria for the game $\mathcal{G}_{\textrm{SV}}$. 

\paragraph{Allocation rule $\mathcal{A}_{\textrm{prob}}$:} We use a probabilistic allocation rule that generalizes the model proposed in~\citet{maehara2015budget}. We define a probabilistic allocation rule called $\mathcal{A}_{\textrm{prob}}$ that allocates ranks to the $k$ admins as follows. 
    \begin{itemize}
        \item \textit{Activation probability:} When an admin $i$ allocates budget to a subject slot, we define his probability of winning that subject slot at a particular rank (determined by the order in which admins play) as 
    \begin{equation}
        \mathrm{PA}_a(\mathbf{x}_a, t) := 1 - \left(1 - p\right)^{\mathbf{x}_a(t)},
    \end{equation}
    where $p$ is a parameter governing how relevant admin $a$ is for a subject slot $t$.\footnote{In this first analysis, we take $p$ to be the same for all admins and subjects; future work can model a different $p_a$ for each admin $a$, noting that the equilibrium becomes more complicated to characterize in closed form.} We use interchangeably the notation $\mathbf{x}_a(t)$ or $\mathbf{x}_{at}$ for the budget that treatment administrator $a$ allocates to slot $t$. We note that the relevance parameter $p$ cannot depend on the specific subject drawn at subject slot $t$, since the admin does not know \textit{a priori} which subject is realized. Thus, we can think of $p$ as an expected value over the population distribution.
    If a treatment administrator does not bid on a slot, he does not get a sample from that slot (at any rank).~\looseness=-1
    \item \textit{Random ordering:} Then, we must define a way for advertisers to compete over the subject slots. We propose the following procedure for administrators to get allocated ranks to show their content to each subject slot: for each subject slot $t \in [n]$ and for each rank $r \leq k$, set $\mathcal{S}_{k - r}$ the set of administrators who have not been allocated a rank $r' < r$ yet for this subject slot $t$.
    Then, draw a permutation of the elements of $\mathcal{S}_{k - r}$, $\sigma \in \mathcal{P}(\mathcal{S}_{k - r})$, uniformly at random (where $\mathcal{P}(S)$ denotes the set of all permutations of elements of a set $S$). Now, according to the order that the permutation $\sigma$ defined, each administrator will try to win rank $r$ for subject slot $t$, based on the budget they allocated to this subject slot (note that an administrator allocates budget to a subject, not to each rank). See Algorithm~\ref{alg:winranks} for a formal description.
    \end{itemize}

\begin{algorithm}[htb]
\caption{Winning ranks for a subject slot  $t$ through the probabilistic allocation rule $\mathcal{A}_{\textrm{prob}}$.}
\label{alg:winranks}
\begin{algorithmic}
\STATE Denote $\mathcal{S}_{k -1 } = [k]$ the set of all administrators. 
\FOR{rank $r \in [k]$:}  
    \STATE Draw a random permutation $\sigma \in \mathcal{P}(\mathcal{S}_{k - r})$
    \FOR{$i \in [k]$:}  
        \STATE Admin $\sigma(i)$ tries to win subject $t$ at rank $r$ with probability $\mathrm{PA}_{\sigma(i)}(x_{\sigma(i)}, t)$
        \IF{success}
            \STATE Define $\mathcal{S}_{k - r - 1} = \mathcal{S}_{k - r} \backslash \sigma(i)$
            \STATE break and move on to the next rank
        \ENDIF
    \ENDFOR
\ENDFOR
\end{algorithmic}
\end{algorithm}

We characterize a set of Nash equilibria for the case ${B^{(a)} \leq n}$ in the following lemma, thus proving Theorem~\ref{prop:negame}. The results easily generalize to $B^{(a)} > n$, by considering allocations in which treatment administrators split their budget uniformly over subjects. 

\begin{lemma}
     An allocation $\textbf{x}$ with the following properties is always a Nash equilibrium for $\mathcal{G}_{\textrm{SV}}$:   
    \begin{itemize}
        \item Admins prefer to split their budget uniformly over subjects: $\mathbf{x}_{a}(t) \in \{ 0,1\}, \forall a \in [k], t \in [n]$; 
        \item Admins prefer to spend all their budget: \[\sum\limits_{t = 1}^{n} \mathbf{x}_a(t) = B^{(a)}, \forall a \in [k] \] 
        \item Admins prefer to minimize competition with each other: \[\left| \sum\limits_{a = 1}^{k} \mathbf{x}_a(t) - \sum\limits_{a = 1}^{k} \mathbf{x}_a(t') \right| \leq 1, \forall t,t' \in [n]\] 
        This means that an admin will prioritize spending budget on the subjects with the least bids.
    \end{itemize}
\label{lemma:nash_kgen-c0}
\end{lemma}

The proof for Lemma~\ref{lemma:nash_kgen-c0} is detailed in the Appendix and relies on the following properties for an allocation $\textbf{x}$ as described: an admin $a$ (1) cannot increase their utility through permuting his bids
(re-arranging his allocation $\mathbf{x}_a$ of budget units over subjects); and (2) prefers to split his budget in units of $1$ over subjects rather than aggregating it over fewer subjects. The intuition behind proving these properties relies on the form of the utility function $f_a(\cdot)$, as $\mathrm{PA}_a(\cdot)$ is component-wise concave and increasing: an admin gains more from uniform bidding since it has a chance of winning a first rank over multiple subjects, rather than increasing his bid on a single subject. Although increasing the bid on a single subject increases his chance to win the first rank, it does not increase it by much in comparison. Moreover, we note that an allocation in which an admin has unused budget cannot be an equilibrium, as the admin can increase their sample value by bidding on new subjects, given the probabilistic nature of the allocation rule. The nature of this equilibrium is quite subtle: it’s not clear from the start whether the equilibrium is always the equal bids strategy, or to split the subject slots (biding high on a few subject slots and zero on the rest, thus splitting the subjects between different admins). We conjecture that the allocations $\textbf{x}$ as described in Lemma~\ref{lemma:nash_kgen-c0} are the only Nash equilibria that can occur when the relevance probability $p$ is the same for all admins:~\looseness=-1

\begin{conjecture}
    The only pure Nash equilibria for the game $\mathcal{G}_{\textrm{SV}}$ are allocations $\mathbf{x}$ that satisfy the properties stated in Lemma~\ref{lemma:nash_kgen-c0}, when the activation probability $p_a := p$ is the same for all admins $a \in [k]$.
    \label{conj:onlynashis01}
\end{conjecture}

We can formally show this property for two admins, $k = 2$, and any number of subjects $n$: 

\begin{proposition}
    The only pure Nash equilibria for the game $\mathcal{G}_{\textrm{SV}}$ for two admins ($k = 2$) are allocations $\mathbf{x}$ that satisfy the properties stated in Lemma~\ref{lemma:nash_kgen-c0}, when the activation probability $p_a := p$ is the same for all admins $a \in [k]$.
    \label{prop:onlynashis01_k2}
\end{proposition}

    \section{Discussion and future directions}
\label{sec7:discussion}

While our analysis uses the probabilistic allocation rule $\mathcal{A}_{\textrm{prob}}$, it is not limited to it. As the approximation proof sketch suggests, a main ingredient is that the inherent randomness in the allocation rule renders the sample value objective $S$ close to its mean, allowing for concentration inequalities to apply and extend to the estimation error objective. Albeit the random ordering of the agents, the allocation rule has the property that higher bids increase the probability of winning higher ranks, due to the sequential rank allocation. The Nash equilibria in the sample value optimization game bear a noteworthy resemblance to~\citet{feldman2007budget}'s randomized solution of the associated optimization problem under a generalized second-price auction. In their setting, the authors show that the optimization problem induced by the utility function of agents with position-based value over subjects is intractable, yet randomized strategies find nearly optimal solutions. Thus, future directions could analyze our objectives with other allocation rules similar to those used in generalized second-price or first-price auctions~\cite{vickrey1961counterspeculation}.~\looseness=-1

Another promising future avenue of research includes studying the quality of Nash equilibria within the class of estimation error games, in terms of bounding the Price of Anarchy (PoA)~\cite{koutsoupias2009worst}. In the $\mathcal{G}_{\textrm{SV}}$ game, this becomes more challenging to study as it is not necessarily a potential game like its simpler version in~\citet{maehara2015budget}, although with the added novelty that we can fully characterize a Nash equilibrium. Furthermore, it would be interesting to study the distribution of utility across admins through the lens of inequality, perhaps through related solution concepts such as Nash social welfare, often employed to study fairness issues in resource allocation games~\cite{branzei2017nash}.~\looseness=-1
    \newpage
\section*{Acknowledgements}
 We thank Florian Dorner for his insightful feedback on the manuscript. We are also grateful to Claire Vernade, Aaron Schein, Kevin Jamieson, Lalit Jain, Robert Nowak, and Scott Sievert for helpful discussions on earlier versions of the project. Finally, we thank the anonymous reviewers for their helpful feedback on the paper. This project has received funding from the Max Planck ETH Center for Learning Systems (CLS).

    \bibliography{main}
    \bibliographystyle{icml2024}
    \newpage
    \appendix
    \section{Appendix}

\subsection{Figure of data-generating process in Section~\ref{sec:causalinference}}
\begin{figure*}[h!]
    \centering
    \includegraphics[width=0.9\textwidth]{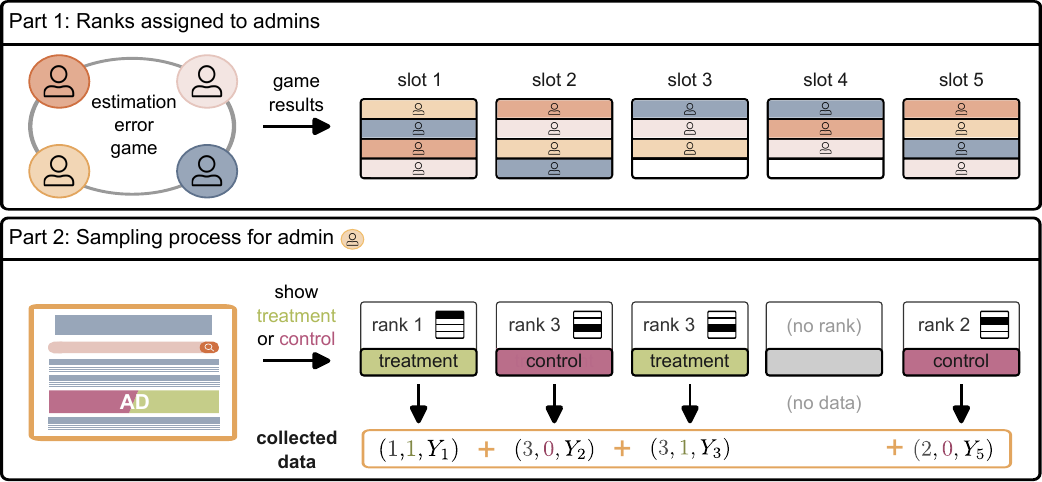}
    \caption{Illustration of the data-generating process for an admin. In Part $1$, the admins compete with each other to obtain different ranks for subject slots. After placing the bids, a rank allocation rule is applied. Each admin receives a rank for all slots for which they have bid. For example, the admin depicted in yellow received rank $1$ for subject slots $1$, rank $3$ for subject slots $2$ and $3$, no rank for subject slot $4$ (which happens if an admin does not bid on a slot), and rank $2$ for subject slot $5$. In Part $2$, individuals are sampled at the allocated ranks, with treatment and control assigned in a randomized manner. The collected data for the admin depicted in yellow consists of tuples $(r_i, T_i, Y_i)$ for all individuals, where $r_i$ represents the rank, $T_i$ the binary treatment-control variable, and $Y_i$ the outcome.}
    \label{fig:causalinference}
\end{figure*}

We sketch the data-generating process for an admin in Figure~\ref{fig:causalinference}.

\newpage

\subsection{Proofs for Section~\ref{sec3:setup} results}
\label{sec6:approxproof}

\begin{proof}[Proof of Theorem~\ref{thm:approxthm}]

We assume that every admin $a$ can only bid up to some constant $C$ units of its budget $B^{(a)}$ on any particular subject slot $t$. This assumption is motivated by throttling behavior on online advertising markets~\cite{charles2013budget} or advertising channel constraints in media~\cite{maehara2015budget}. We also assume that the budget of any admin is at most the number of subject slots: $B^{(a)} \leq n$. We note that this assumption is not restrictive, as results  hold easily for larger budgets.~\looseness=-1

We aim to show that for $\mathbf{x}$ a Nash equilibrium for the $\mathcal{G}_{\textrm{SV}}$ game, $\mathbf{x}$ is an $(\epsilon,\eta)$-approximate Nash equilibrium for the $\mathcal{G}_{\textrm{MSE}}$ game. For ease of notation, denote the sample value by a variable $S$:
\begin{equation}
    S := \sum\limits_{r = 1}^k n_r \cdot \alpha_r^2.
\label{eq:defineS}
\end{equation}
The sample value $S$ is determined by the allocation $\mathbf{x}$. Note that $S$ is a positive random variable, as as the variables $(n_r)_r$ counts of the rank allocation $\overline{r}$, $n_r = \sum_i \mathbbm{1}\{r_i = r \}$ and the variables $(\alpha_r)_r$ are nonnegative.
The rank allocation $\overline{r}$ is sampled from the distribution given by the allocation rule. We aim to show the following result, as a stepping stone in proving the existence of an $\epsilon$-approximate Nash equilibrium for the game $\mathcal{G}_{\textrm{MSE}}$:

\begin{proposition}
    For allocations $\mathbf{x} \in \mathcal{D}$ in which all admins use all their budget, $\sum\limits_{t = 1}^n \mathbf{x}_a(t) = B^{(a)}$, $\forall a \in [k]$, with high probability $S \in \left[(1 - \epsilon) \cdot \mathbb{E}[S], (1 + \epsilon) \cdot \mathbb{E}[S] \right]$, with $\epsilon > 0$.
    \label{prop:fullallocationsconcentrate}
\end{proposition}

\begin{proof}[Proof of Proposition~\ref{prop:fullallocationsconcentrate}]

The intuition is that the sample value concentrates around its expectation with high probability given that it satisfies a bounded differences property, as further shown. Our aim is to show that with high probability 

\begin{equation}
    \begin{gathered}
        S \in \left[(1 - \epsilon) \cdot \mathbb{E}[S], (1 + \epsilon) \cdot \mathbb{E}[S] \right] \Leftrightarrow \\ 
        (1 + \epsilon) \cdot \mathbb{E}[S] \geq S \geq (1 - \epsilon) \cdot \mathbb{E}[S] \Leftrightarrow \\ 
        \epsilon \cdot \mathbb{E}[S] \geq S - \mathbb{E}[S] \geq -\epsilon \cdot \mathbb{E}[S] \Leftrightarrow \\ 
        | S - \mathbb{E}[S] | \leq \epsilon\cdot \mathbb{E}[S]
    \end{gathered}
    \label{eq:event}
\end{equation}

In order to show this, we upper bound the probability $\mathbb{P}\left( | S - \mathbb{E}[S] | \geq \epsilon\cdot \mathbb{E}[S] \right)$ using McDiarmid's inequality by writing the random variable $S$ as the output of a function that satisfies the bounded differences property. 
An admin $a$ only obtains samples from subject slots on which they bid, i.e. if they bid $0$ budget on a subject slot, they do not get \textit{any} rank for that subject slot, as per the definition of the allocation rule.
Given the assumption that an admin can only bid up to some constant $C$ on each subject, there are at least $B^{(a)} / C$ slots where the admin gets any rank; hence, there are at least $B^{(a)} / C$ of the $r_i$ variables. We denote by $m$ the number of the subject slots on which an admin gets ranks, noting that $B^{(a)} / C \leq m \leq B^{(a)}$. 

For each sample profile $(n_r)_r$, we can write $S$ as the output of a function of the random variables $(r_i)_i$, $g: [k]^{m} \rightarrow \mathbb{R}$, with $g(r_1, \cdots, r_{m}) = \sum\limits_{r = 1}^k \sum\limits_{i = 1}^m \mathbbm{1}(r_i = r) \cdot \alpha_r^2$ (in this notation, $n_r$ becomes $n_r = \sum\limits_{i = 1}^m \mathbbm{1}(r_i = r)$).

We next show that the function $g$ satisfies the bounded differences property in the variables $(r_i)_i: \exists $ constants $c_i, \forall i \in [n]$ such that $\forall i \in [n]$ and $r_i \in [k]$, the following property is satisfied: 

\begin{equation}
    \sup\limits_{r_i' \in [k]} | g(r_1,\cdots, r_i, \cdots, r_m) - g(r_1,\cdots, r_i', \cdots, r_m) | \leq c_i    
\end{equation}

To show this, we denote 

\begin{equation}
    \begin{gathered}
        g(r_1,\cdots,r_i,\cdots,r_m) = \sum\limits_{r = 1}^k n_r \cdot \alpha_r^2, \\
        g(r_1,\cdots,r_i',\cdots,r_m) = \sum\limits_{r = 1}^k n_r' \cdot \alpha_r^2,
    \end{gathered}
\end{equation}
where $n_{r_i}' = n_{r_i} -1 $, $n_{r_i'}' = n_{r_i'} + 1$, and for $r \neq r_i, r_i'$, $n_r = n_r'$, since changing one variable $r_i$ to $r_i'$ only changes two counts of the ranks, $n_{r_i}$ and $n_{r_i'}$, leaving the rest unchanged. Thus, we get that 

\begin{equation}
    \begin{gathered}
        \sup\limits_{r_i' \in [k]} | g(r_1,\cdots,r_i,\cdots,r_m)  - g(r_1,\cdots, r_i', \cdots, r_m) | = \sup\limits_{r_i' \in [k]} \left| \alpha_{r_i}^2 - \alpha_{r_i'}^2 \right| \leq 1,
    \end{gathered}
\end{equation}
since $\alpha_r \in [0,1], \forall r \in [k]$. Thus, the function $g$ satisfies the bounded differences property with right-handside constants equal to $1$ for all $i \in [n]$. From McDiarmid's inequality~\cite{mcdiarmid1989method}, we get that for independent random variables $R_i \in [k], \forall i \in [m]$,

\begin{equation}
    \begin{gathered} 
    \mathbb{P}\left( |g(R_1,\cdots,R_m) - \mathbb{E}\left[ g(R_1,\cdots,R_m)\right] | \geq \eta \right) \leq 2 \exp \left( - \frac{2\eta^2}{m} \right) \Leftrightarrow \\ 
    \mathbb{P}\left( | S - \mathbb{E}\left[ S \right] | \geq \eta \right) \leq 2 \exp \left( - \frac{2\eta^2}{m} \right)
    \end{gathered}    
    \label{eq:mcdiarmideta}
\end{equation}
Thus, the probability that the even in equation~\ref{eq:event} happens is bounded below: 

\begin{equation}   
\mathbb{P}\left( | S - \mathbb{E}[S] | \leq \epsilon\cdot \mathbb{E}[S] \right) \geq 1 - 2 \exp \left( - \frac{2\epsilon^2 \mathbb{E}[S]^2}{m} \right)
\label{eq:final}
\end{equation}
As $\epsilon \rightarrow 0$, the right hand side of equation~\ref{eq:final} approaches $1.$

\end{proof}

Next, we will use the properties of the allocation rule $\mathcal{A}_{\textrm{prob}}$ that allow us to bound $\mathbb{E}[S]$:

\begin{proposition}
    If $B^{(a)} \leq n $ for an admin $a$, then under the allocation rule $\mathcal{A}_{\textrm{prob}}$ and a Nash equilibrium $\mathbf{x}$, $\mathbb{E}\left[ \sum\limits_{r = 1}^k n_r \cdot \alpha_r^2 \right] = \Omega( B^{(a)} / k)$.
        \label{prop:probofwinningranksbounded}
\end{proposition} 

\begin{corollary}    
    If $B^{(a)} \leq n$ for an admin $a$, then under the allocation rule $\mathcal{A}_{\textrm{prob}}$ and an allocation $\mathbf{x}$ with $\sum\limits_{t = 1}^n \mathbf{x}_a(t) = B^{(a)}$, $\mathbb{E}\left[ \sum\limits_{r = 1}^k n_r \cdot \alpha_r^2 \right] = \Omega( B^{(a)} / C \cdot k)$.
        \label{cor:probofwinningranksbounded}
\end{corollary}

\begin{proof}[Proof of Proposition~\ref{prop:probofwinningranksbounded}]
    First, we note by linearity of expectation and the fact that all variables are non-negative that $\mathbb{E}\left[ \sum\limits_{r = 1}^{k} n_r \cdot \alpha_r^2 \right] = \sum\limits_{r = 1}^{k} \mathbb{E}\left[ n_r\right] \cdot \alpha_r^2 \geq \mathbb{E}\left[ n_1\right] \cdot \alpha_1^2 = \mathbb{E}\left[ n_1\right]$, as we set $\alpha_1 = 1$. The allocation rule $\mathcal{A}_{\textrm{prob}}$ gives us a closed form formula for $\mathbb{E}\left[ n_1\right] $, 

    \begin{equation}
        \mathbb{E}\left[ n_1\right] = \sum\limits_{t = 1}^{n}  \mathrm{PA}_a(\mathbf{x}_a, t) \cdot \frac{1}{k!} \sum\limits_{\sigma \in \mathcal{P}([k])} \mathrm{PA}_a(\mathbf{x}_a,t) \cdot \prod\limits_{j <_{\sigma} a} \left(1 - \mathrm{PA}_j(\mathbf{x}_j,t)\right),
        \label{eq:expofn1}
    \end{equation}
    where $\mathcal{P}(K)$ denotes the set of permutation of elements of a set $K$. We note that each summand in equation~\eqref{eq:expofn1} is the probability that admin $a$ wins rank $1$ for a subject slot $t$, which we write as 

    \begin{equation}
        p_{a,t}(\mathbf{x}) = \mathrm{PA}_a(\mathbf{x}_a, t) \cdot \frac{1}{k!} \sum\limits_{\sigma \in \mathcal{P}([k])} \mathrm{PA}_a(\mathbf{x}_a,t) \cdot \prod\limits_{j <_{\sigma} a} \left(1 - \mathrm{PA}_j(\mathbf{x}_j,t)\right),
    \end{equation}
    thus giving 
    
    \begin{equation}
        \mathbb{E}\left[n_1 \right] = \sum\limits_{t = 1}^{n} p_{a,t}(\mathbf{x})
    \end{equation}

As $p_{a,t}$ is essentially averaging over all possible orderings of the admins, we note that 

\begin{equation}
    p_{a,t}(\mathbf{x}) = \sum\limits_{s = 0}^{k - 1} p_{a,t}(\mathbf{x} | \mbox{ admin } a \mbox{ appears after } s \mbox{ admins in } \sigma) \cdot \mathbb{P}\left(\mbox{ admin } a \mbox{ appears after } s \mbox{ admins in }  \sigma \right),
\end{equation}
where $\sigma$ is a random ordering of the admins. Thus, 

\begin{equation}
    p_{a,t}(\mathbf{x}) \geq p_{a,t}(\mathbf{x} | \mbox{ admin } a \mbox{ appears first in } \sigma) \cdot \mathbb{P}\left(\mbox{ admin } a \mbox{ appears first in }  \sigma \right)
\end{equation}

Since $\sigma$ is a random ordering of the admins, the probability that admin $a$ is the first one in $\sigma$ is simply $1/k$, whereas the probability that admin $a$ wins rank $1$ for subject slot $1$ conditioned on the fact that it the first one to play is just $\mathrm{PA}_a(\mathbf{x}_a,t)= 1 - (1-p)^{\mathbf{x}_a(t)}$ given the allocation rule $\mathcal{A}_{\textrm{prob}}$. Thus, we obtain 

\begin{equation}
    \mathbb{E}\left[n_1 \right] \geq \sum\limits_{t = 1}^n \frac{\mathrm{PA}_a(\mathbf{x}_a,t)}{k}
\end{equation}

Under a Nash equilibrium $\mathbf{x}$, we know from Lemma~\ref{lemma:nash_kgen-c0} that admin $a$ has exhausted their budget $B^{(a)}$ and has bid only $1$s and $0$s. Thus,

\begin{equation}
    \mathbb{E}\left[n_1 \right] \geq \sum\limits_{\substack{t \in [n], \\\mathbf{x}_a(t) = 1}} \frac{\mathrm{PA}_a(\mathbf{x}_a,t)}{k} + \sum\limits_{\substack{t \in [n], \\ \mathbf{x}_a(t) = 0}} \frac{\mathrm{PA}_a(\mathbf{x}_a,t)}{k}
    \label{eq:expn1fornash}
\end{equation}

Since $\mathrm{PA}_a(\mathbf{x}_a,t)= 1 - (1-p)^{\mathbf{x}_a(t)}$, we get that $\mathrm{PA}_a(\mathbf{x}_a,t)= p$ for $\mathbf{x}_a(t) = 1$ and $\mathrm{PA}_a(\mathbf{x}_a,t)= 0$ for $\mathbf{x}_a(t) = 0$, thus giving 

\begin{equation}
\begin{gathered}    
    \mathbb{E}\left[n_1 \right] \geq \sum\limits_{\substack{t \in [n], \\\mathbf{x}_a(t) = 1}} \frac{\mathrm{PA}_a(\mathbf{x}_a,t)}{k}  \Rightarrow \\
    \mathbb{E}\left[n_1 \right] \geq \frac{B^{(a)} \cdot p}{k}
\end{gathered}
\label{eq:expn1fornash2}
\end{equation}

Since $p$ is defined as a constant with respect to all other model parameters, we get that $\mathbb{E}\left[n_1\right] = \Omega\left(\frac{B^{(a)}}{k}\right)$. We conclude that $\mathbb{E}\left[ \sum\limits_{r = 1}^k n_r \cdot \alpha_r^2 \right] = \Omega( B^{(a)} / k)$, which follows from the simple property that if a function $f$ is $\Omega(n)$, then for function $g$ for which $g(x) \geq f(x), \forall x$ in the domain of $f$ and $g$, $g$ also grows as $\Omega(n)$. We note that there may be other allocations rule that satisfy this property, and thus our analysis is not contingent on the specific choice of the allocation rule, as long as it preserves this property.
\end{proof}

\begin{proof}[Proof for Corollary~\ref{cor:probofwinningranksbounded}]
Corollary~\ref{cor:probofwinningranksbounded} follows easily from the above proof with a minimal modification of equations~\eqref{eq:expn1fornash}--\eqref{eq:expn1fornash2}, knowing that an admin can bid at most $C$ budget units on every subject slot and they have exhausted their budget:

\begin{equation}
    \mathbb{E}\left[n_1 \right] \geq \sum\limits_{\substack{t \in [n], \\\mathbf{x}_a(t) \geq 1}} \frac{\mathrm{PA}_a(\mathbf{x}_a,t)}{k} + \sum\limits_{\substack{t \in [n], \\ \mathbf{x}_a(t) = 0}} \frac{\mathrm{PA}_a(\mathbf{x}_a,t)}{k}
    \label{eq:expn1forfullbid}
\end{equation}

Since $\mathrm{PA}_a(\mathbf{x}_a,t)= 1 - (1-p)^{\mathbf{x}_a(t)}$, we get that $\mathrm{PA}_a(\mathbf{x}_a,t) \geq p$ for $\mathbf{x}_a(t) \geq 1$ and $\mathrm{PA}_a(\mathbf{x}_a,t)= 0$ for $\mathbf{x}_a(t) = 0$, thus giving 

\begin{equation}
\begin{gathered}    
    \mathbb{E}\left[n_1 \right] \geq \sum\limits_{\substack{t \in [n], \\\mathbf{x}_a(t) \geq 1}} \frac{\mathrm{PA}_a(\mathbf{x}_a,t)}{k}  \Rightarrow \\
    \mathbb{E}\left[n_1 \right] \geq \frac{B^{(a)} \cdot p}{C \cdot k}
\end{gathered}
\label{eq:expn1forfullbid2}
\end{equation}
under the assumption that $B^{(a)} \leq n$. Just as before, we get that $\mathbb{E}\left[n_1\right] = \Omega\left(\frac{B^{(a)}}{C \cdot k}\right)$ and conclude that $\mathbb{E}\left[ \sum\limits_{r = 1}^k n_r \cdot \alpha_r^2 \right] = \Omega( B^{(a)} /( C \cdot k))$.

\end{proof}

Corollary~\ref{cor:probofwinningranksbounded} simply states that the expectation of the variable $S$ grows at at least the rate of $B^{(a)} / k$ (since $C$ is a constant) when an admin uses their entire budget, whether in a Nash equilibrium or not (intuitively, subject to using all budget, any permutation of how the budget is allocated across subject slots, the expectation of $S$ does not get `too small'). Take a constant $c > 0$ such that
\begin{equation}
    \mathbb{E}[S] \geq c \cdot \frac{B^{(a)}}{C \cdot k}.
    \label{eq:ctsforESbkbound}
\end{equation}

Since $B^{(a)} / C\leq m \leq B^{(a)}$, we also know that there exists a constant $c' > 0$ such that $\mathbb{E}[S] \geq c' \cdot \frac{m}{k}$. We take $\eta$ in equation~\eqref{eq:mcdiarmideta} to be equal to $\eta = c_0 \cdot \sqrt{m}$, with $c_0$ to be determined. Then, equation~\eqref{eq:mcdiarmideta} becomes

\begin{equation}
    \begin{gathered}
    \mathbb{P}\left( | S - \mathbb{E}\left[ S \right] | \leq \eta \right) \geq 1 - 2 \exp \left( - \frac{2\eta^2}{m} \right) \Leftrightarrow \\
    \mathbb{P}\left( | S - \mathbb{E}\left[ S \right] | \leq \sqrt{m} \cdot c_0 \right) \geq 1 - 2 \exp \left( - 2c_0^2 \right)
    \end{gathered}
\end{equation}

Taking $c_0 = \epsilon \cdot c' \cdot \frac{\sqrt{m}}{k}$, we get from equation~\eqref{eq:ctsforESbkbound}: 

\begin{equation}
    \begin{gathered}
            \mathbb{P}\left( | S - \mathbb{E}\left[ S \right] | \leq \epsilon \cdot c' \cdot \frac{m}{k} \right) \geq 1 - 2 \exp \left( - 2\epsilon^2 c'^2 m / k^2 \right) \Rightarrow \\ 
            \mathbb{P}\left( | S - \mathbb{E}\left[ S \right] | \leq \epsilon \cdot \mathbb{E}[S] \right) \geq 1 - 2 \exp \left( - 2\epsilon^2 c'^2 m / k^2 \right) 
    \end{gathered}
    \label{eq:mcdiarmidsfinal}
\end{equation}

The right hand side tends to $1$ when the exponential vanishes, i.e. in the regime where $B^{(a)} = \omega(k^2)$. Various modeling assumptions may satisfy this requirement, for example when the budget and the number of administrators grow sublinearly in $n$, with $B^{(a)} = \Omega(n^{b})$ and $k = \mathcal{O}(n^{b/2 - \delta})$, for $b \in (0,1)$ and $\delta \in (0, b/2)$. 


A simple corollary follows from Proposition~\ref{prop:fullallocationsconcentrate}:

\begin{corollary}
    For allocations $\mathbf{x} \in \mathcal{D}$ in which all admins use all their budget, $\sum\limits_{t = 1}^n \mathbf{x}_a(t) = B^{(a)}$, $\forall a \in [k]$, with high probability $ \frac{1}{S} \in \left[(1 - \epsilon) \cdot \frac{1}{\mathbb{E}[S]}, (1 + \epsilon)\cdot  \frac{1}{\mathbb{E}[S]}\right]$, with $\epsilon > 0$.
    \label{cor:fmseconcentrates0}
\end{corollary}

\begin{proof}
    The proof follows easily by constructing $\epsilon$ and using Proposition~\ref{prop:fullallocationsconcentrate}. Since we know that with high probability, $S \geq (1 - \epsilon) \cdot \mathbb{E}[S]$ for an allocation $\mathbf{x}$ and $S$ is positive, 

    \begin{equation}        
    \begin{gathered}
        S \geq (1 - \epsilon) \cdot \mathbb{E}[S] \Rightarrow \frac{1}{S} \leq \frac{1 }{(1 - \epsilon) \cdot \mathbb{E}[S]}, \\
        S \leq (1 + \epsilon) \cdot \mathbb{E}[S] \Rightarrow \frac{1}{S} \geq \frac{1 }{(1 + \epsilon) \cdot \mathbb{E}[S]} 
    \end{gathered}
        \end{equation}

    Set $\frac{1}{1 - \epsilon} = 1 + \epsilon_0 \Rightarrow \epsilon = \frac{\epsilon_0}{1 + \epsilon_0} \Rightarrow \frac{1}{1 + \epsilon} = 1 - \frac{\epsilon_0}{1 + 2 \epsilon_0}$. Choosing $\epsilon' = \max \left(\epsilon_0, \frac{\epsilon_0}{1 + 2\epsilon_0}\right)$, we get that 

    \begin{equation}
        \mathbb{P}\left( \left| \frac{1}{S} - \frac{1}{\mathbb{E}[S]} \right| \geq \epsilon' \cdot \frac{1}{\mathbb{E}[S]} \right) = \mathbb{P}\left( | S - \mathbb{E}[S]| \geq \epsilon \cdot \mathbb{E}[S] \right) \geq 2 \exp \left( - 2\epsilon^2 c'^2 m / k^2 \right),
    \end{equation}
    concluding our proof. For $m/k^2 \rightarrow \infty$, the right hand side vanishes. 
\end{proof}

We return to the proof of Theorem~\ref{thm:approxthm}.
We denote by $f_a^{\textrm{MSE}}$ the utility function of the game $\mathcal{G}_{\textrm{MSE}}$ and by $f_a^{\textrm{SV}}$ the utility function of the game $\mathcal{G}_{\textrm{SV}}$, for each admin $a$.

Note that $f_a^{\textrm{SV}}$ is a non-negative function on its entire domain for all admins $a$, whereas $f_a^{\textrm{MSE}}$ is a non-positive function on its entire domain for all admins $a$. Note that $f_a^{\textrm{MSE}}(\mathbf{x}) = - \mathbb{E} \left[ \frac{1}{\sum\limits_{r = 1}^k n_r \cdot \alpha_r^2 } \right] = - \mathbb{E} \left[ \frac{1}{S} \right]$ and $f_a^{\textrm{SV}}(\mathbf{x}) = \mathbb{E} \left[ \sum\limits_{r = 1}^k n_r \cdot \alpha_r^2\right] = \mathbb{E} \left[ S\right]$, by definition of the random variable $S$ from equation~\eqref{eq:defineS}.

    According to Proposition~\ref{prop:fullallocationsconcentrate}, we know that $S$ concentrates around its expectation. For $\epsilon > 0$, denote the event that $S \in \left[ (1 - \epsilon) \cdot \mathbb{E}[S], (1 + \epsilon) \cdot \mathbb{E}[S] \right]$ by $\mathcal{E}$ for ease of notation. Thus, 

    \begin{equation}
        \mathbb{E}\left[ \frac{1}{S}\right] = \mathbb{E}\left[ \frac{1}{S} | \mathcal{E}\right]\cdot \mathbb{P}\left[\mathcal{E}\right] + \mathbb{E}\left[ \frac{1}{S} | \mathcal{E}^c\right]\cdot \mathbb{P}\left[ \mathcal{E}^c\right], 
    \end{equation}
where $A^c$ denotes the complement of the event $A$. We know that, conditioned on $\mathcal{E}$, $\frac{1}{S}$ is concentrated around $\frac{1}{\mathbb{E}[S]}$, whereas conditioned on $\mathcal{E}^c$, expectation of $\frac{1}{S}$ is bounded by $1$ given that the error estimation objective outputs $\min \left(\frac{1}{S}, 1\right)$. Denote by $\eta$ the probability of $\mathcal{E}$ (from Proposition~\ref{prop:fullallocationsconcentrate}, $\eta = 2e^{-2\epsilon^2c'^2 m/k^2}$). Thus, we obtain 

\begin{equation}
    \begin{gathered}
        \mathbb{E}\left[ \frac{1}{S}\right] \leq \mathbb{E}\left[ \frac{1}{S} | \mathcal{E}\right] + 1 \cdot \eta \leq ( 1 + \epsilon) \cdot \frac{1}{\mathbb{E}[S]} + \eta \Rightarrow \\ 
        f_a^{\textrm{MSE}}(\mathbf{x}) \geq (1 + \epsilon) \left( - \frac{1}{f_a^{\textrm{SV}}(\mathbf{x})}\right) - \eta
        \end{gathered}
        \label{eq:approxfmse_fvs}
\end{equation}
which holds for all allocations $\mathbf{x}$ such that all admins have allocated all their budget, under the assumption that an admin can bid at most $C$ budget units on any subject slot.

Furthermore, applying Jensen's inequality on the function $f(x) = 1/x$, we get that for all allocations $\mathbf{x}_a$

\begin{equation}
\begin{gathered}
    \mathbb{E} \left[ \frac{1}{\sum\limits_{r = 1}^k n_r \cdot \alpha_r^2}\right] \geq \frac{1}{\mathbb{E}\left[ \sum\limits_{r = 1}^k n_r \cdot \alpha_r^2 \right]} \Leftrightarrow \\ 
    - \mathbb{E} \left[ \frac{1}{\sum\limits_{r = 1}^k n_r \cdot \alpha_r^2}\right] \leq - \frac{1}{\mathbb{E}\left[ \sum\limits_{r = 1}^k n_r \cdot \alpha_r^2 \right]} \Leftrightarrow \\
     f_a^{\textrm{MSE}}(\mathbf{x}_a, \mathbf{x}_{-a}) \leq - \frac{1}{f_a^{\textrm{SV}}(\mathbf{x}_a, \mathbf{x}_{-a})} 
    \end{gathered}
    \label{eq:fullalloc3}
\end{equation}

Now, take $\mathbf{x}$ to be a Nash equilibrium for the game, and we will show that $\mathbf{x}$ has to be an $(\epsilon,\eta)$-approximate Nash equilibrium for the game $\mathcal{G}_{\textrm{MSE}}$. Thus, we would like to show that $\forall \mathbf{x}_a' \in \mathcal{D}_a$, 

\begin{equation}
    \begin{gathered}
        f_a^{\textrm{SV}}(\mathbf{x}_a, \mathbf{x}_{-a}) \geq f_a^{\textrm{SV}}(\mathbf{x}_a', \mathbf{x}_{-a}) \Rightarrow \\ 
    f_a^{\textrm{MSE}}(\mathbf{x}_a, \mathbf{x}_{-a}) \geq (1 + \epsilon) \cdot f_a^{\textrm{MSE}}(\mathbf{x}_a', \mathbf{x}_{-a}) - \eta
    \end{gathered}
    \label{eq:epsetapproxcondition}
\end{equation}
with $\epsilon,\eta > 0, \forall a \in [k]$. 

We will first show that the above statement holds for all $\mathbf{x}_a' \in \mathcal{D}_a$ such that $\sum\limits_{t = 1}^n \mathbf{x}_a'(t) = B^{(a)}$ (in other words, admin $a$ has exhausted their budget with strategy $\mathbf{x}_a'$). Given that $\mathbf{x}$ is a Nash equilibrium for $\mathcal{G}_{\textrm{SV}}$, we get

\begin{equation}
    \begin{gathered}
            f_a^{\textrm{SV}}(\mathbf{x}_a, \mathbf{x}_{-a}) \geq f_a^{\textrm{SV}}(\mathbf{x}_a', \mathbf{x}_{-a}) \Rightarrow \\ 
    - \frac{1}{ f_a^{\textrm{SV}}(\mathbf{x}_a, \mathbf{x}_{-a})} \geq - \frac{1}{ f_a^{\textrm{SV}}(\mathbf{x}_a', \mathbf{x}_{-a})} \Rightarrow \\ 
    f_a^{\textrm{MSE}}(\mathbf{x}_a, \mathbf{x}_{-a}) \geq (1 + \epsilon) \left(- \frac{1}{ f_a^{\textrm{SV}}(\mathbf{x}_a, \mathbf{x}_{-a})}\right) - \delta \geq \\
    \geq (1 + \epsilon) \left(- \frac{1}{ f_a^{\textrm{SV}}(\mathbf{x}_a', \mathbf{x}_{-a})}\right) - \delta  \geq (1 + \epsilon) \cdot f_a^{\textrm{MSE}}(\mathbf{x}_a', \mathbf{x}_{-a})- \eta
    \end{gathered}
    \label{eq:epsetafullalloc}
\end{equation}
where the last inequality follows from Jensen's inequality, as described above. Since $\epsilon,\eta > 0$, we get the $(\epsilon,\eta)$-approximation condition. Thus, we are left to show inequality~\eqref{eq:epsetapproxcondition} for allocations $\mathbf{x}_a'$ that may not exhaust the entire budget $B^{(a)}$. For each such allocation, create another allocation $\mathbf{x}_a^{\textrm{full}}$ by using the rest of the budget randomly across subject slots that had no bid on $\mathbf{x}_a'$. Thus, $\mathbf{x}_a^{\textrm{full}}$ satisfies the conditions necessary for Proposition~\ref{prop:fullallocationsconcentrate}. We note that the utility function $f_a^{\textrm{SV}}$ can only increase from bidding a unit of budget or more on additional subjects which had a previous bid of $0$ budget, since $f_a^{\textrm{SV}}$ is a summation of the utility coming from all subjects $t\in[n]$ with all terms being non-negative, and $\mathrm{PA}_a(\mathbf{x}_a,t) = 0$ if $\mathbf{x}_a(t) = 0$ and $\mathrm{PA}_a(\mathbf{x}_a,t) \geq p > 0$ if $\mathbf{x}_a(t) \geq 1$. Thus, we get 

\begin{equation}
    f_a^{\textrm{SV}}(\mathbf{x}_a^{\textrm{full}}, \mathbf{x}_{-a}) \geq f_a^{\textrm{SV}}(\mathbf{x}_a', \mathbf{x}_{-a}) \Leftrightarrow - \frac{1}{f_a^{\textrm{SV}}(\mathbf{x}_a^{\textrm{full}},\mathbf{x}_{-a})} \geq - \frac{1}{f_a^{\textrm{SV}}(\mathbf{x}_a', \mathbf{x}_{-a})} 
    \label{eq:fullalloc2}
\end{equation}

From equations~\eqref{eq:epsetafullalloc}--\ref{eq:fullalloc3} and using the Jensen inequality again, we obtain that 

\begin{equation}
    f_a^{\textrm{MSE}}(\mathbf{x}_a^{\textrm{full}}, \mathbf{x}_{-a}) \geq  (1 + \epsilon) \left( - \frac{1}{f_a^{\textrm{SV}}(\mathbf{x}_a^{\textrm{full}}, \mathbf{x}_{-a})} \right) - \eta \geq (1 + \epsilon) \left( - \frac{1}{f_a^{\textrm{SV}}(\mathbf{x}_a', \mathbf{x}_{-a})} \right) - \eta  \geq (1 + \epsilon) \cdot f_a^{\textrm{MSE}}(\mathbf{x}_a', \mathbf{x}_{-a}) - \eta,
\end{equation}
thus proving that a Nash equilibrium for the game $\mathcal{G}_{\textrm{SV}}$ is an $(\epsilon,\eta)$-approximate Nash equilibrium for the game $\mathcal{G}_{\textrm{MSE}}$. We note that $\epsilon$ and $\eta$ are 
\begin{equation}
    \epsilon = \mathcal{O}\left(\frac{k}{\sqrt{B}}\right)
    \quad\text{and}\quad
    \eta = \mathcal{O}\left(\exp(-B/k^2)\right),
\end{equation}
    where $B=\min_a B^{(a)}$ and the maximum bid per subject slot is bounded. These come from equation~\eqref{eq:mcdiarmidsfinal}, in which the concentration bounds links the concentration `closeness' $\epsilon$ with the probability of concentration $\eta$. 
    
\end{proof}

\newpage 

\subsection{Proofs for Section~\ref{sec:causalinference} results}
We define a population distribution $\cX$ on the subjects, which are fully described by the random variables $(\rank,T,Y)$.
Individuals are drawn independently for each slot, according to a fixed rank allocation $\overline{r}$. In particular, we sample the $n_r$ individuals from their slots at rank $r$ from the do-interventional distribution $\cX^{\textrm{do}(\rank = r)}$. 

We highlight that the outcomes $Y_i$ for $i \in [n^{(a)}]$ are \emph{not} identically distributed. Their distribution differs according to their associated ranks $r_i$. The ranks $r_i$ are assigned by the rank allocation $\overline{r}$, and therefore, deterministic in $i$. They merely indicate from which do-interventional distribution the individual was drawn.
We denote the joint distribution of the outcomes and treatment assignment variables $Z_i=(Y_i,T_i), i\in[n^{(a)}]$ by $P$.~\looseness=-1 

We explicitly define the set of considered model instances $\mathcal{M}$. For a given sequence $(\alpha_r)_{r=1,\ldots,k}$, we set
\begin{equation}
    \mathcal{M} = \left\{\left(c_{i,0},c_{i,1}^{(r_i)}\right)_{i\in[n^{(a)}]} : |\tau| \leq 1,  \tau_r = \alpha_r \cdot \tau \right\},
\end{equation}
with the quantities $\tau = \frac{1}{n^{(a)}}  \sum_{i=1}^{n^{(a)}} \frac{c_{i,1}^{(r_i)}}{\alpha_{r_i}}$ and 
\[
    \tau_r = 
    \begin{cases}
        \frac{1}{n_r} \sum\limits_{i:r_i=r} c_{i,1}^{(r)} &\text{for} \ n_r\geq 1\\
        \alpha_r \cdot \tau & \text{for} \ n_r =  0
    \end{cases}
\]
for $r\in[k]$.\footnote{We note that when there are no datapoints at rank $r$, we cannot identify the effect directly, but we can identify it from data at other ranks since the structural parameters $(\alpha_r)_r$ are known. }

Moreover, note that we also identify the causal effect of interest in equation~\eqref{eq:estimand_rankr} through the potential outcomes model as
\begin{equation*}
    \tau = \frac{1}{n^{(a)}}\sum_{i=1}^{n^{(a)}} \frac{c_{i,1}^{(r_i)}}{\alpha_{r_i}}.
\end{equation*}

\begin{proof}[Proof of Theorem~\ref{thm:minimax-lowerbound}]

We follow the proof of Theorem 2 in~\citet{cortez2022exploiting} and apply Le Cam's method~\cite{lecam1973minimax} to obtain the minimax bound. For the computation, we make use of the independence between individuals, and the fact that we can sort them according to their assigned ranks.

We use Le Cam's method~\cite{lecam1973minimax} to obtain the proof of Theorem~\ref{thm:minimax-lowerbound}.
\begin{lemma}[Le Cam's method]
    Let $\mathcal{M}$ be a set of model instances. Then, we have for any two instances $M_1, M_2 \in \mathcal{M}$ such that $|\tau(M_1) - \tau(M_2) | \geq 2\delta$ with $\delta>0$:
    \begin{equation}
        \inf_{\widehat{\tau}} \sup_{M \in \mathcal{M}} \E\left[(\widehat{\tau}-\tau)^2\right]
        \geq \frac{\delta^2}{2} \left(1- \sqrt{\frac{D_{KL}(P_1||P_2)}{2}}\right),
    \end{equation}
    where $P_1$ and $P_2$ denote the population distribution with respect to $M_1$ and $M_2$, respectively, and $D_{KL}(\cdot||\cdot)$ denotes the Kullback-Leibler divergence.
    \label{lemma:lecam}
\end{lemma}

    We begin the proof by constructing instances $M_1$ and $M_2$. We set $c_{i,0}=0$ and $c_{i,1}^{r_i} = \delta \cdot \alpha_{r_i}$ for $M_1$ and $c_{i,0}=0$ and $c_{i,1}^{r_i} = - \delta \cdot \alpha_{r_i}$ for $M_2$. Note that $\alpha_1 = 1$ by definition. Hence, we get
    \begin{align}
        \tau(M_1) &= \frac{1}{n^{(a)}} \sum_{i=1}^{n^{(a)}} \frac{c_{i,1}^{(r_i)}}{\alpha_{r_i}} = \frac{1}{n^{(a)}} \sum_{i=1}^{n^{(a)}} \frac{\delta \cdot \alpha_{r_i}}{\alpha_{r_i}} = \delta\\
        \tau(M_2) &= \cdots = - \delta
    \end{align}
    So, we have $|\tau(M_1)-\tau(M_2)|\geq 2\delta$. Thus, the instances $M_1$ and $M_2$ satisfy the coverage condition in Lemma~\ref{lemma:lecam}.

    Next, we compute $D_{KL}(P_1||P_2)$ with $P_1$ and $P_2$ are the joint distribution of $(Y_i,T_i), i\in[n^{(a)}]$ for the instances $M_1$ and $M_2$. We will choose a $\delta$ to upper bound $D_{KL}(P_1||P_2)$ by $1/2$. As all samples are independently drawn, the joint distributions $P_1$ and $P_2$ factorize as:
    \begin{align*}
        P_1(\{(Y_i,T_i)\}_{i=1}^{n^{(a)}}) &= \prod_{i=1}^{n^{(a)}} P_1(Y_i,T_i)\\
        P_2(\{(Y_i,T_i)\}_{i=1}^{n^{(a)}}) &= \prod_{i=1}^{n^{(a)}} P_2(Y_i,T_i).
    \end{align*}
    Note that the outcomes $Y_i$ and $Y_j$ are differently distributed when the subjects $i$ and $j$ are sampled from different ranks, $r_i \neq r_j$. Hence, we have in general that $P_1(Y_i,T_i) \neq P_1(Y_j,T_j)$ and $P_2(Y_i,T_i) \neq P_2(Y_j,T_j)$ for $i\neq j$. We obtain that
    \begin{equation*}
        D_{KL}(P_1||P_2) = \E_{M_1}\left[\log\left(\frac{p_1(\{(Y_i,T_i)\}_{i=1}^{n^{(a)}})}{p_2(\{(Y_i,T_i)\}_{i=1}^{n^{(a)}})}\right)\right] = \sum_{i=1}^{n^{(a)}} \E_{M_1}\left[\log\left(\frac{p_1(Y_i,T_i)}{p_2(Y_i,T_i)}\right)\right] 
    \end{equation*}
    with $p_1$ and $p_2$ the densities of $P_1$ and $P_2$, respectively.
    We sort the samples according to their allocated ranks and apply the Law of Total Expectation,
    \begin{align}
        D_{KL}(P_1||P_2) =\sum_{r=1}^k \sum\limits_{j:r_j=r}& \E_{M_1}\left[\log\left(\frac{p_1(Y_j,T_j)}{p_2(Y_j,T_j)}\right)\right] \nonumber \\
        = \sum_{r=1}^k \sum\limits_{j:r_j=r}& P_1(T_j=1) \cdot \E_{M_1}\left[\log\left(\frac{p_1(Y_j,T_j)}{p_2(Y_j,T_j)}\right)| T_j=1\right] \nonumber\\
        &+  P_1(T_j=0) \cdot \E_{M_1}\left[\log\left(\frac{p_1(Y_j,T_j)}{p_2(Y_j,T_j)}\right)|T_j=0\right]
        \label{eq:appendix-minimax-proof-treatment-zero}
    \end{align}
    Focusing on samples drawn from a fixed rank $r$, the outcomes $Y_j$ conditional on the treatment $T_j$, are distributed as:
    \begin{equation}
        Y_j |T_j \sim N\left(c_{i,1}^{(r)}\cdot T_j, \sigma^2\right)
    \end{equation}
    as $Y_j = c_{j,0} + c_{j,1}^{(r)}\cdot T_j + \varepsilon_j$ with $c_{j,0} = 0$ and $\varepsilon_j\sim N(0,\sigma^2)$.
    Hence, we have for $M_1$
    \begin{equation*}
        Y_j |T_j \sim N(\delta \cdot \alpha_{r}\cdot T_j, \sigma^2)
    \end{equation*}
    and for $M_2$
    \begin{equation*}
        Y_j |T_j \sim N(-\delta \cdot \alpha_{r}\cdot T_j, \sigma^2).
    \end{equation*}
    We have $Y_j |(T_j = 0) \sim N(0, \sigma^2)$ for both instances $M_1$ and $M_2$. 
    Therefore, the second term in equation~\eqref{eq:appendix-minimax-proof-treatment-zero} simplifies to zero. We remain with
    \begin{align*}
        D_{KL}(P_1||P_2)
        &= \sum_{r=1}^k \sum\limits_{j:r_j=r} P_1(T_j=1) \cdot \E_{M_1}\left[\log\left(\frac{p_1(Y_j,T_j)}{p_2(Y_j,T_j)}\right)| T_j=1\right]\\
        &= \sum_{r=1}^k \sum\limits_{j:r_j=r} P_1(T_j=1) \cdot \E_{M_1}\left[\log\left(\frac{\varphi_{\delta \cdot \alpha_{r},\sigma^2}(Y_j,T_j)}{\varphi_{-\delta \cdot \alpha_{r},\sigma^2}(Y_j,T_j)}\right)| T_j=1\right]\\
    \end{align*}
    where $\varphi_{\mu,\sigma^2}$ denotes the density function of $N(\mu,\sigma^2)$. Note that the difference in distribution across ranks $r$ is now made explicit in the densities, e.g., $\varphi_{\delta \cdot \alpha_{r},\sigma^2}$
    for model instance $M_1$.
    We input the explicit formula for the normal density and use that $T_j \sim \textrm{Ber}(q_j)$, for $q_j \in [q, 1-q]$,
    \begin{align}
        D_{KL}(P_1||P_2)
        &= \sum_{r=1}^k \sum\limits_{j:r_j=r} q_j \cdot \E_{M_1}\left[-\frac{1}{2\sigma^2} \left((Y_i-\delta\cdot\alpha_r)^2-(Y_i+\delta\cdot\alpha_r)^2\right)| T_j=1\right]\\
        &= \sum_{r=1}^k \sum\limits_{j:r_j=r} q_j\cdot \frac{2}{\sigma^2} \cdot \delta\cdot\alpha_r \cdot \E_{M_1}\left[ Y_j| T_j=1\right]\\
        &= \sum_{r=1}^k \sum\limits_{j:r_j=r} q_j\cdot \frac{2}{\sigma^2} \cdot \delta\cdot\alpha_r \cdot \left(0+\delta\cdot\alpha_r+\E[\varepsilon_j]\right)\\
        &= \sum_{r=1}^k \sum\limits_{j:r_j=r} q_j \cdot \frac{2}{\sigma^2} \cdot\delta^2\cdot\alpha_r^2\\
        &\leq \delta^2\cdot (1-q)\cdot \frac{2}{\sigma^2} \cdot \sum_{r=1}^k n_r \cdot\alpha_r^2.
    \end{align}
    
    We can bound $D_{KL}(P_1||P_2)$ by $1/2$ if we set
    \begin{equation*}
        \delta^2 = \frac{\sigma^2}{4\cdot (1-q)\cdot \sum_{r=1}^k n_r \cdot \alpha_r^2}.
    \end{equation*}
    Then, Le Cam's method gives,
    \begin{align}
        \inf_{\widehat{\tau}} \sup_{M \in \mathcal{M}} \E\left[(\widehat{\tau}-\tau)^2\right]
        &\geq \frac{\delta^2}{2} \left(1- \sqrt{\frac{D_{KL}(P_1||P_2)}{2}}\right)\\
        &= \frac{\sigma^2}{16\cdot (1-q)\cdot \sum_{r=1}^k n_r \cdot \alpha_r^2} 
    \end{align}
    \begin{figure}[t!]
    \centering
    \includegraphics[width=0.45\textwidth]{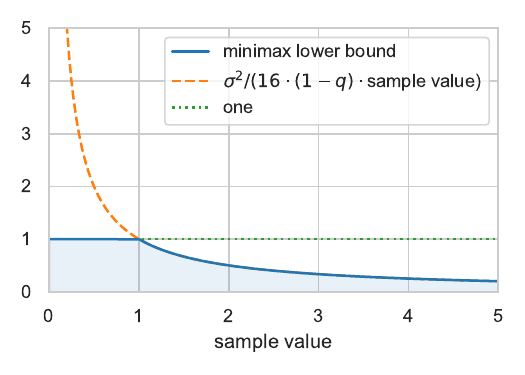}
    \caption{Example of minimax lower bound with $q=0.5$ and $\sigma^2=4$.}
    \label{fig:minimax-lowerbound}
\end{figure}
Finally, note that the sample value $\sum_r n_r \cdot \alpha_r^2$ is lower bounded by one if the admin attains at least one slot at rank one, i.e., $n_1=1$. However, it might happen that no slot is won at rank one (in fact, it may happen that $n_r = 0$ for some other rank $r$ too). In this case, the sample value is sometimes lower than one, and as it decreases, the inverse of $\left( \sum_{r = 1}^k n_r \cdot \alpha_r^2 \right)^{-1}$ may explode. However, the minimum error is still upper bounded by $1$ since the admin is better off using a constant estimator, as the effect is bounded, $|\tau|\leq1$. Therefore, the minimax lower bound is the minimum of the error achieved using the samples and the error obtained by a constant estimator. See Figure~\ref{fig:minimax-lowerbound} for an illustration.

\end{proof}

\begin{proof}[Proof of Proposition~\ref{prop:taurisunbiased}]
   We note that $\tau_r = \frac{1}{n_r} \sum\limits_{i:r_i=r} c_{i,1}^{(r)}$ from the potential outcomes model.

Then, we compute the expectation of our estimator as: 

\begin{equation}
\begin{gathered}    
    \mathbb{E}(\widehat{\tau}_{r}) = \mathbb{E}\left[\frac{1}{n_r} \sum\limits_{i:r_i = r}   \left( \frac{Y_i \mathbbm{1}\{ T_i = 1 \}}{\mathbb{P}(T_i = 1 )} - \frac{Y_i \mathbbm{1}\{ T_i = 0 \}}{\mathbb{P}(T_i = 0 )}\right) \right] \\ 
    = \frac{1}{n_r} \sum\limits_{i:r_i = r}  \mathbb{E}\left[ \frac{Y_i \mathbbm{1}\{ T_i = 1 \}}{\mathbb{P}(T_i = 1 )} - \frac{Y_i \mathbbm{1}\{ T_i = 0 \}}{\mathbb{P}(T_i = 0 )} \right] 
    \end{gathered}
    \label{eq:exp-ht-exclusive}
\end{equation}

For every $i$, note that, by linearity of expectation and the potential outcomes model for subjects $i$ at rank $r_i=r$,

\begin{equation}
    \begin{gathered}
        \mathbb{E}\left[\frac{Y_i \mathbbm{1}\{ T_i = 1 \}}{\mathbb{P}(T_i = 1 )} - \frac{Y_i \mathbbm{1}\{ T_i = 0 \}}{\mathbb{P}(T_i = 0 )} \right] \\ =
        c_{i,0} \cdot \mathbb{E}\left[ \frac{\mathbbm{1}\{ T_i = 1 \}}{\mathbb{P}(T_i = 1 )} - \frac{\mathbbm{1}\{ T_i = 0 \}}{\mathbb{P}(T_i = 0 )}  \right] \\
        + c_{i,1}^{(r)} \cdot \mathbb{E}\left[T_i \cdot \left(\frac{\mathbbm{1}\{ T_i = 1 \}}{\mathbb{P}(T_i = 1 )} - \frac{\mathbbm{1}\{ T_i = 0 \}}{\mathbb{P}(T_i = 0 )} \right)  \right]\\
        + \mathbb{E}[ \varepsilon_i ] \cdot \mathbb{E}\left[ \frac{\mathbbm{1}\{ T_i = 1 \}}{\mathbb{P}(T_i = 1 )} - \frac{\mathbbm{1}\{ T_i = 0 \}}{\mathbb{P}(T_i = 0 )}   \right] \\ =  
        c_{i,0} \cdot \mathbb{E}\left[ \frac{\mathbbm{1}\{ T_i = 1 \}}{\mathbb{P}(T_i = 1 )} - \frac{\mathbbm{1}\{ T_i = 0 \}}{\mathbb{P}(T_i = 0 )}   \right] \\
        + c_{i,1}^{(r)} \cdot \mathbb{E}\left[ T_i \cdot \left(\frac{\mathbbm{1}\{ T_i = 1 \}}{\mathbb{P}(T_i = 1 )} - \frac{\mathbbm{1}\{ T_i = 0 \}}{\mathbb{P}(T_i = 0 )} \right)  \right]\\
    \end{gathered}
    \label{eq:exp-linearity-exclusive}
\end{equation}

Assuming that the actual treatment $T_i$ is administered through a Bernoulli randomized design, ${T_i  } \sim \textrm{Ber}(q_i)$, for $q_i \in [q,1-q]$, we get that the first term of equation~\eqref{eq:exp-linearity-exclusive} reduces to $0$ and the second one to $c_{i,1}^{(r)}$. To see this briefly, note that the first term is equal to

\begin{equation}
    \begin{gathered}
        c_{i,0} \cdot \mathbb{E}\left[\frac{\mathbbm{1}\{ T_i = 1\}}{\mathbb{P}(T_i = 1)]} - \frac{\mathbbm{1}\{ T_i = 0\}}{\mathbb{P}(T_i = 0)}  \right] \\
        = c_{i,0}  \cdot \mathbb{E}\left[  \frac{\mathbbm{1}\{ T_i = 1\}}{\mathbb{P}(T_i = 1)]} - \frac{\mathbbm{1}\{ T_i = 0\}}{\mathbb{P}(T_i = 0)} \right] \\ 
        = c_{i,0} \cdot \left( \frac{\mathbb{P}(T_i = 1)}{\mathbb{P}(T_i = 1)} - \frac{\mathbb{P}(T_i = 0)}{\mathbb{P}(T_i = 0)}\right) = 0
    \end{gathered}
    \label{eq:exp-linearity-exclusive-t1}
\end{equation}
using the law of total expectation. Similarly, the second term is equal to

\begin{equation}
    \begin{gathered}
        c_{i,1}^{(r)} \cdot \mathbb{E}\left[ T_i \cdot \left(\frac{\mathbbm{1}\{ T_i = 1 \}}{\mathbb{P}(T_i = 1 )} - \frac{\mathbbm{1}\{ T_i = 0 \}}{\mathbb{P}(T_i = 0 )} \right)  \right] \\
        = c_{i,1}^{(r)}  \cdot \mathbb{E} \left[ T_i \cdot \left(\frac{\mathbbm{1}\{ T_i = 1 \}}{\mathbb{P}(T_i = 1 )} - \frac{\mathbbm{1}\{ T_i = 0   \}}{\mathbb{P}(T_i = 0 )} \right)  \right] \\
        = c_{i,1}^{(r)} \cdot \left( \mathbb{E} \left[ T_i \cdot \frac{\mathbbm{1}\{ T_i = 1\}}{\mathbb{P}(T_i = 1  )}   \right]
        - \mathbb{E} \left[T_i \cdot \frac{\mathbbm{1}\{ T_i = 0\}}{\mathbb{P}(T_i = 0 )}   \right] \right) \\
        = c_{i,1}^{(r)} \cdot \left( \mathbb{E} \left[ \mathbbm{1}\{ T_i = 1\} \cdot \frac{\mathbbm{1}\{ T_i = 1\}}{\mathbb{P}(T_i = 1  )}   \right]
        - \mathbb{E} \left[\mathbbm{1}\{ T_i = 1\} \cdot \frac{\mathbbm{1}\{ T_i = 0\}}{\mathbb{P}(T_i = 0 )}   \right] \right)\\
        = c_{i,1}^{(r)}  \cdot \left( \mathbb{E} \left[ \frac{\mathbbm{1}\{ T_i = 1  \}}{\mathbb{P}(T_i = 1  )} \right]
        - 0 \right)\\
        =  c_{i,1}^{(r)} \cdot \frac{\mathbb{P}(T_i = 1 )}{\mathbb{P}(T_i = 1 )} =  c_{i,1}^{(r)}
    \end{gathered}
    \label{eq:exp-linearity-exclusive-t2}
\end{equation}
as $T_i = \mathbbm{1}\{ T_i = 1\}$ by definition, and we know that $\mathbbm{1}\{ T_i = 1\} \cdot\mathbbm{1}\{ T_i = 1\} = \mathbbm{1}\{ T_i = 1\}$ and $\mathbbm{1}\{ T_i = 1\} \cdot\mathbbm{1}\{ T_i = 0\} =0$. \\

Coming back to the sum over all subjects $i$ at rank $r$, we get that 

\begin{equation}
    \mathbb{E}(\widehat{\tau}_{r}) = \frac{1}{n_r} \cdot \sum\limits_{i:r_i = r}  c_{i,1}^{(r)} = \tau_r
\end{equation}

\end{proof}

\begin{proof}[Proof for Proposition~\ref{prop-hterror-ranked}]
    
    Denote by 

    \begin{equation}
    \begin{gathered}
        \widehat{\tau}_{r,d} = \frac{1}{n_r} \cdot \sum\limits_{i:r_i=r} \frac{Y_i \mathbbm{1}\{T_i = d  \}}{\mathbb{P}(T_i = d )},
    \end{gathered}
    \label{eq:horvitzthompson_estimator_rankr-1}
\end{equation}
for $d \in \{0, 1\}$. Then, we have $\widehat{\tau}_r = \widehat{\tau}_{r,1} - \widehat{\tau}_{r,0}$. We begin by computing the variance of $\widehat{\tau}_{r,d}$ for $d \in \{0,1 \}$.

\paragraph{Step 1:}

\begin{equation}
    \begin{gathered}
        \textrm{Var}(\widehat{\tau}_{r,d}) = \frac{1}{n_r^2} \sum\limits_{i:r_i = r} \sum\limits_{j:r_j = r}  \frac{\textrm{Cov}(Y_i\mathbbm{1}\{T_i=d  \}, Y_j\mathbbm{1}\{T_j=d  \})}{\mathbb{P}(T_i = d )\mathbb{P}(T_j = d )}
    \end{gathered}
\end{equation}

Let's first take the case $d = 1$. 
Case 1: For $i = j$, 

\begin{equation}
\begin{gathered}
    \textrm{Cov}(Y_i\mathbbm{1}\{T_i=1  \},Y_j\mathbbm{1}\{T_j=1 \}) 
    = \textrm{Var}(Y_i\mathbbm{1}\{T_i=1  \})
\end{gathered}
\label{eq:cov_computation_rankr}
\end{equation}

Note that 

\begin{equation}
    \begin{gathered}
        \textrm{Var}(Y_i \mathbbm{1}\{T_i = 1  \}) = \textrm{Var}(\varepsilon_i) \cdot \textrm{Var}(\mathbbm{1}\{T_i = 1  \}) + \textrm{Var}(\varepsilon_i) \cdot \mathbb{E}(\mathbbm{1}\{T_i = 1  \})^2 \\
        +  \mathbb{E}(Y_i|T_i)^2 \cdot \textrm{Var}(\mathbbm{1}\{T_i = 1  \}) \\ 
        \leq \sigma^2 \cdot q_i(1 - q_i) + \sigma^2 \cdot q_i^2 + \left(c_{i,0}+c_{i,1}^{(r)}+\mathbb{E}(\varepsilon_i)\right)^2 \cdot q_i(1-q_i) \\
        = \sigma^2 \cdot q_i +Y_{\textrm{max}}^2 \cdot q_i(1-q_i)
    \end{gathered}
\end{equation}
where the last inequality holds under the bounded effect assumption from equation~\ref{assumption:effectboundrank}.

Case 2: For $i \neq j$, under the Bernoulli design, treatment assignments are independent and so no treatment affects both units $i$ and $j$. Hence, $\textrm{Cov}(Y_i\mathbbm{1}\{T_i=1  \},Y_j\mathbbm{1}\{T_j=1  \}) = 0$. \\

Thus, 
\begin{equation}
    \begin{gathered}
        \textrm{Var}(\widehat{\tau}_{r,1}) \leq \frac{1}{n_r^2} \sum\limits_{i:r_i = r}  \frac{\sigma^2 \cdot q_i +Y_{\textrm{max}}^2 \cdot q_i(1-q_i)}{t^2} \\ 
        = \frac{1}{n_r^2} \sum\limits_{i:r_i = r}  \frac{\sigma^2 +Y_{\textrm{max}}^2 \cdot (1-q_i)}{q_i} \\
        \leq \frac{1}{n_r \cdot q} \left( \sigma^2 + Y_{\textrm{max}}^2 \cdot (1 -q)\right),
    \end{gathered}
    \label{eq:bounding_var_taur0}
\end{equation}
given that the second equation is a decreasing function in $q_i$ and $q_i \in [q,1-q]$. In a similar way we obtain a bound for $d = 0$: 

\begin{equation}
    \begin{gathered}
        \textrm{Var}(\widehat{\tau}_{r,0}) \leq \frac{1}{n_r \cdot (1-q)} \left( \sigma^2 + Y_{\textrm{max}}^2 \cdot q\right) 
    \end{gathered}
    \label{eq:bounding_var_taur1}
\end{equation}

\paragraph{Step 2:}
Next, we bound the covariance between $\widehat{\tau}_{r,0}$ and $\widehat{\tau}_{r,1}$:

\begin{equation}
    \begin{gathered}
        \textrm{Cov}(\widehat{\tau}_{r,1},\widehat{\tau}_{r,0}) = \mathbb{E}[\widehat{\tau}_{r,1} \cdot \widehat{\tau}_{r,0}] - \mathbb{E}[\widehat{\tau}_{r,1}] \cdot \mathbb{E}[\widehat{\tau}_{r,0}]
    \end{gathered}
\end{equation}

A simple calculation shows that

\begin{equation}
    \mathbb{E}[\widehat{\tau}_{r,t}] = \frac{1}{n_r} \sum\limits_{i:r_i = r}  Y_i(T_i = t)
\label{eq:single_exp_r}
\end{equation}
and 

\begin{equation}
\begin{gathered}
    \mathbb{E}[\widehat{\tau}_{r,1} \cdot \widehat{\tau}_{r,0}] = \mathbb{E}\left[ \frac{1}{n_r^2} \left( \sum\limits_{i:r_i = r}   \frac{Y_i \mathbbm{1}(T_i = 1 )}{\mathbb{P}(T_i = 1 )} \right) \cdot \left( \sum\limits_{i:r_i = r}  \frac{Y_i \mathbbm{1}(T_i = 0 )}{\mathbb{P}(T_i = 0 )} \right)\right] \\
\end{gathered}
\label{eq:joint_exp_r}
\end{equation}

Using equations~\eqref{eq:single_exp_r} and~\eqref{eq:joint_exp_r} in the covariance expression and simplifying, we get

\begin{equation}
    \begin{gathered}
    \textrm{Cov}(\widehat{\tau}_{r,1},\widehat{\tau}_{r,0}) = \frac{1}{n_r^2} \left( \sum\limits_{i\neq j} \frac{\mathbb{P}(T_i = 1 \& T_j = 0)}{\mathbb{P}(T_i = 1 )\mathbb{P}(T_j = 0)}
    \cdot Y_i(T_i = 1) Y_j(T_j = 0) - \sum\limits_{i:r_i = r} Y_i(T_i = 1) \sum\limits_{i:r_i = r} Y_i(T_i = 0)\right)\\
    = \frac{1}{n_r^2} \left( \sum\limits_{i\neq j} Y_i(T_i = 1)Y_j(T_j = 0) -  \sum\limits_{i,j} Y_i(T_i = 1)Y_j(T_j = 0) \right)\\
    = - \frac{1}{n_r^2}\sum\limits_{i:r_i = r} Y_i(T_i = 1, )Y_i(T_i = 0)
    \end{gathered}
\end{equation}

Given the bounded effect assumption from equation~\ref{assumption:effectboundrank}, we can bound
\begin{equation}
    \begin{gathered}
    -\textrm{Cov}(\widehat{\tau}_{r,1},\widehat{\tau}_{r,0}) \leq \frac{Y_{\textrm{max}}^2}{n_r}
    \end{gathered}
    \label{eq:covbounds_r}
\end{equation}

\paragraph{Step 3:}
Putting equations~\eqref{eq:bounding_var_taur0},~\eqref{eq:bounding_var_taur1}, and~\eqref{eq:covbounds_r} together, we get

\begin{equation}
    \begin{gathered}
        \textrm{Var}(\widehat{\tau}_r) = \textrm{Var}(\widehat{\tau}_{r,1}) + \textrm{Var}(\widehat{\tau}_{r,0}) - 2 \cdot \textrm{Cov}(\widehat{\tau}_{r,1},\widehat{\tau}_{r,0}) \Rightarrow \\ 
        \textrm{Var}(\widehat{\tau}_r) \leq \frac{1}{n_r} \left[ \frac{\sigma^2}{q(1-q)} + Y_{\textrm{max}}^2 \left( \frac{(1-q)^2 + q^2}{q(1-q)} + 2 \right)\right] 
    \end{gathered}
\end{equation}

\end{proof}

\newpage
\subsection{A discussion on estimating the discount factors}

When the discount factors are not known, an admin is put in the following position: should he discard the (potentially useful) data at lower ranks and just use the data at rank $1$ for estimating the treatment effect, or should he try to engage the lower rank data by using part of it to estimate the discount factors and part of it for estimating the treatment? At a first glance, the latter might sound appealing, especially if there was significantly more data at lower ranks than at rank $1$. However, this latter option does not actually decrease the estimation error, even in idealized cases.
We showcase this claim in an ideal situation below, where the lower rank data exhibits no error in the $\tau_r$ estimator (i.e. $\mathbb{E} [\left(\tau_r - \hat{\tau}_r \right)^2] = 0$ for $r > 1$). This result uses a na\"{i}ve estimator for the discount parameters $(\alpha_r)_r$, leaving the problem of using more complex estimators for future work.~\looseness=-1

To formalize our set-up, we define: 
\begin{itemize}
    \item Scenario (1): an admin uses just the data where their treatment was shown on the first rank, and thus $\tau = \tau_1$. We denote an estimator for this scenario by $\hat{\tau}_{(1)}$, noting that it uses $n_1$ samples. 
    \item Scenario (2): an admin splits the data at all positions, using part of it for estimating the position effect parameters $(\alpha_r)_r$, and then using these estimates together with the rest of the data for estimating the treatment effect $\tau$. We denote an estimator for this scenario by $\hat{\tau}_{(2)}$. 
\end{itemize}

\begin{lemma}
    Assume a na\"{i}ve estimator for the discount factors: $\hat{\alpha}_r = \hat{\tau}_r/ \hat{\tau}_1, \forall r$. For all valid ways of splitting the data between estimating $(\alpha_r)_r$ and estimating $\tau$,~\looseness=-1
    \begin{equation*}\mathbb{E}[(\hat{\tau}_{(1)} - \tau)^2] \leq \mathbb{E}[(\hat{\tau}_{(2)} - \tau)^2]
    \end{equation*}
    \label{prop:jointestimation}
\end{lemma}

\begin{proof}
First, splitting the data in a valid way means using some amount of samples $n'$ from both the rank $1$ data and rank $r$ data to estimate $\alpha_r$ with the na\"{i}ve estimator, and using the remaining (non-overlapping) samples to estimate $\tau$. One must have $n' \leq n_1$ and $n' \leq n_r$. Since we have $k - 1$ such estimation problems, for every rank $r \in [k]$ except the first one, one must use $k - 1$ disjoint sample sets from the rank $1$ data, totalling at most the number of samples at rank $1$ (since we cannot use more data than available). 

In scenario (1), the admin can only use the data at rank $1$ in order to estimate $\tau$ (since it is the treatment effect had the ads been only shown at rank $1$ to all subjects). Without any knowledge of $(\alpha_r)_r$, the admin cannot use the data at rank $2, 3,$ etc to identify the treatment effect through $\tau_{r}$. Thus, $\tau = \tau_1$, which, as we have showed in Theorem~\ref{thm:minimax-lowerbound} and Proposition~\ref{prop-hterror-ranked}, has error $\mathbb{E} \left[ (\tau - \widehat{\tau})^2\right] = \Theta \left( n_1^{-1}\right)$ (recall that $n_1$ is the sample size at position $1$). 

In scenario (2), the admin uses the samples at all positions, relating the effect $\tau$ from data at rank $r$ through the $\widehat{\alpha}_r$ estimate: $\widehat{\tau}^{(r)} = \frac{\widehat{\tau}_r}{\widehat{\alpha}_r}$.
Note that the data used for estimating $\widehat{\alpha}_r$ must be separate from the data used for estimating $\widehat{\tau}^{(r)}$.
Let's say that an admin uses $n_r'$ samples from the $n_r$ samples at each rank for estimating $\widehat{\alpha}_r$. The rest of $n_r - n_r'$ samples will be used to estimate the $r$-estimand. In an ideal case where the estimator at rank $r$ is a perfect estimator, we have $\widehat{\tau}_r = \tau_r$. In this case, we only have to worry about the contribution of the $\widehat{\alpha}_r$ estimates to the estimation error. For example, one can imagine that we have an abundance of data at lower ranks, $n_r >> 1$  for $r > 1$, and thus the estimation error of $\tau_r$ vanishes with increased number of samples. 

Note that $\hat{\tau}_{(2)}$ depends on the values $(n_r')_r$. However, the result holds no matter what these values are, i.e. no matter how many samples the admin uses from each rank.
This is at first unintuitive. The sample size of the (final) estimator in scenario (2), $\sum_r n_r'$, may be much larger than $n_1$, and yet, that does not help in reducing the estimation error. At a closer look, we'll see that the error contributed by the discount factors estimation, $(\widehat{\alpha}_r)_r$, is a dominant term in the error of the treatment effect.

Take as a naive estimator for $\alpha_r$ the ratio of the treatment effects at rank $r$ and rank $1$:
$\widehat{\alpha}_r = \frac{\widehat{\tau}_r}{\widehat{\tau}_1}$, $\forall r > 1$, using $n_r'$ samples from the $n_r$ samples from rank $r$ and $n_r'$ samples from the $n_1$ samples at rank $1$; i.e., we use $2n_r'$ samples to estimate $\alpha_r$, split equally between rank $1$ and rank $r$ data. The remaining $n_r - n_r'$ samples from the data at rank $r$ are kept for contributing to estimating $\tau$. To better keep track of the number of samples used, we denote an estimator $\hat{\theta}$ based on $n$ samples by $\hat{\theta}(n)$.
Then, $\widehat{\alpha}_r = \frac{\widehat{\tau}_r(n_r')}{\widehat{\tau}_1(n_r')}$, $\forall r > 1$. So, $\frac{1}{\widehat{\alpha}_r} = \frac{\widehat{\tau}_1(n_r')}{\widehat{\tau}_r} =  \frac{\widehat{\tau}_1( n_r')}{\tau_r}$.

Using the remaining samples, the $r-$estimand can be estimated through $\frac{\widehat{\tau}_r(n_r - n_r')}{\widehat{\alpha}_r} = \frac{\tau_r}{\widehat{\alpha}_r} = \widehat{\tau}_1(n_r')$. Thus, the $r$-estimand uses the $n_r'$ samples used for the estimator of the discount factors. The $n_r - n_r'$ samples used in the $\tau_r$ estimation do not factor in because we assumed it is a perfect estimator. 

Thus, the available data at all ranks can be used to estimate $\tau$ through the identified effects at each rank, using $n_1 - \sum_r n_r'$ samples for rank $1$, $n_2'$ samples for rank $2$, and so on until $n_k'$ samples for rank $k$. Just as in the main text, $\alpha_1 = 1$, so we can use the entire remaining samples at rank $1$ directly, i.e., $n_1' = 0$.

Theorem~\ref{thm:minimax-lowerbound} computes a lower bound based on the combination of the $r$-estimands. The lower bound obtain is a summation of the errors each of them contributes, namely over the sequence $(n_r \cdot \alpha_r^2)_r$. This connects to the sample value definition from Section~\ref{sec:objdesign}. In our case, the same $r$-estimands contribute a sequence of errors $(n_1 - \sum_r n_r', n_2', \cdots, n_k')$. Applying Theorem~\ref{thm:minimax-lowerbound} with the same proof but this updated sequence, we obtain an error lower bound of $\Omega \left( \left((n_1 -  \sum_r n_r') + n_2' + \cdots n_k'\right)^{-1} \right) = \Omega(n_1^{-1})$.
(Note that $n_1' = 0$ since $\alpha_1 = 1$.)
The constants are, in fact, the same as in the original Theorem~\ref{thm:minimax-lowerbound}. This updated bound does not depend on the amount of samples used to estimate $\alpha_r$, namely, on $(n_r')_r$. Thus, the lower bound remains the same no matter how many or how few samples an admin has used to estimate the discount factors.~\looseness=-1

To see this exemplified, we use the Horvitz-Thompson estimator for $\widehat{\tau}_1$, and define the estimand as a weighted sum of the $r-$estimands: $\widehat{\tau} = \sum_r w_r \cdot \widehat{\tau}^{(r)}$. As $\widehat{\tau}^{(r)}$ are estimated through $\widehat{\tau}_1$ applied on either $n_1 - \sum_r n_r'$ or $n_r'$ samples for each rank, it is still unbiased. We can then apply the optimal weighting argument from the main text that dictates that error achieved by $\widehat{\tau}$ through optimal weighting is $\mathcal{O}\left( \left( \sum_r 1/ \mathrm{Var} \left( \widehat{\tau}^{(r)}\right) \right)^{-1}\right)$. We know from Proposition~\ref{prop-hterror-ranked} that $\mathrm{Var} \left( \widehat{\tau}^{(r)}\right) = C \cdot \frac{1}{n_r}$ when there are $n_r$ data samples for a constant $C$. We thus obtain an upper bound on the treatment effect estimation error of 

\begin{equation}
    \mathcal{O}\left( \left( \sum_r 1/ \mathrm{Var} \left( \widehat{\tau}^{(r)}\right) \right)^{-1}\right) = \mathcal{O}\left( \left( (n_1 - \sum_r n_r') + n_2' + \cdots + n_k' \right)^{-1}\right) = \mathcal{O}\left( n_1^{-1}\right),
\end{equation}
matching the lower bound (up to constants). 

One would have imagined that having more samples used from rank $2,3,$ and so on that contribute to estimating the treatment effect would have helped reduced the treatment effect estimation error. However, even in this idealized scenario where one has perfect estimation of $\tau_r$ at each rank $r$, the contribution of the discount factors error to the treatment effect error trumps using more low-ranked samples. In realistic scenarios where $\widehat{\tau}_r$ is not a perfect estimator for $r> 1$, one would expect that their error contributes to the estimation error in scenario (2). In this case, an admin can only increase their error by attempting to estimate $(\alpha_r)_r$ from an ad campaign, and is better off (up to constants) just using the $n_1$ samples it has from rank $1$, scenario (1). By how much does an admin increases their estimation error is a complex question, depending on the modeling assumptions and estimators chosen for the discount factors. We leave this for future work. Related work has proposed various methods for estimating the discount factors~\cite{Chuklin:2016cm,joachims2017unbiased}, but to our knowledge, none answer this particular question of when is it actually worth doing that to reduce the treatment effect estimation error.~\looseness=-1

\end{proof}

\newpage

\subsection{Proofs for Section~\ref{sec:gametheory} results}

\begin{proof}[Proof for Lemma~\ref{lemma:nash_kgen-c0}]

Our goal is to show that if an allocation $\mathbf{x}$ satisfies the conditions of Lemma~\ref{lemma:nash_kgen-c0}, i.e. $\mathbf{x}_a(t) \in \{0,1\}, \forall a\in [k], t \in [n]$, $\sum_{t = 1}^n x_a(t) = B^{(a)}, \forall a\in[k]$, and $ | \sum_{a = 1}^{k} \mathbf{x}_a(t) - \sum_{a = 1}^{k} \mathbf{x}_a(t')| \leq 1, \forall t,t' \in [n]$, then it is an equilibrium. We show this through two properties: 
\begin{itemize}
    \item \textbf{Property 1:} an admin $a$ cannot increase his utility by permuting the bids in $\mathbf{x}$. 
    \item \textbf{Property 2:} an admin $a$ cannot increase his utility by aggregating his budget (merging bids from multiple subjects in $\mathbf{x}$).~\looseness=-1
\end{itemize}

In other words, we show that an admin cannot increase his utility by deviating from $\mathbf{x}$, by analyzing the two possible types of deviations: permutations or aggregation of bids.

\paragraph{Notation:} We introduce simplifying notation in order to prove the two aforementioned properties. As the utility function of an admin is the expected sample value, by linearity of expectation, we can write 

    \begin{equation}
        f_a(\mathbf{x}) = \sum\limits_{r = 1}^{K} \mathbb{E} \left[ n_r \right] \cdot \alpha_r^2
    \end{equation}
    
    Given our allocation rule, we can write the utility function in combinatorial form, by finding the closed for of $\mathbb{E}\left[n_r\right], \forall r \in [k]$, 

        \begin{equation}
    \begin{gathered}
        f_a(\mathbf{x}) = \sum_{t = 1}^n \sum_{r = 1}^{k} \alpha_r^2 \cdot \frac{1}{{k - 1 \choose k - r}} \cdot \sum_{s \in \mathcal{S}_{k - r}^{(a)}} \left[ \frac{1}{(k -r + 1)!} \sum_{\sigma \in \mathcal{P}(s \cup \{ a\})} \mathrm{PA}_a(\mathbf{x}_a,t) \cdot \prod_{j <_{\sigma} a} (1 - \mathrm{PA}_j(\mathbf{x}_j, t))\right]   \\
        \cdot \prod_{\overline{r} < r} \left( 1 - \frac{1}{{k - 1 \choose k - \overline{r}}} \cdot \sum_{\overline{s} \in S_{k - \overline{r}}^{(a)}} \frac{1}{(k - \overline{r}  + 1)!} \sum_{\overline{\sigma} \in \overline{s} \cup \{a\}} \mathrm{PA}_a(\mathbf{x}_a,t) \cdot \prod_{\overline{j} <_{\overline{\sigma}} a} (1 - \mathrm{PA}_j(\mathbf{x}_j,t))  \right),
    \end{gathered}  
    \label{eq:utilityfct}
    \end{equation}
    where $\mathcal{S}_{k - r}^{(a)}$ is the set of all choices of $k - r$ elements of $[k] \backslash \{ a\}$ and $j <_{\sigma} a$ means that element $j$ appeared before element $a$ in the permutation $\sigma$. The summand is the probability of admin $a$ to win rank $r$ for slot $t$, by summing over all possible permutations at each rank and possible remaining admins (who have not yet won a rank). The probability of admin $a$ to win a rank is the activation probability times the probability that no one else before $a$ has won, and times the probability that $a$ has not won a previous rank. 

To make things a bit more readable, take the summand for a particular subject $t$ outside of the factor $\alpha_r^2$ and denote it by 

\begin{equation}
    g_{r,a,t}(\mathbf{x}_a, \mathbf{x}_{-a}) := \mathrm{PA}_a(\mathbf{x}_a,t)  \cdot \frac{1}{{k - 1 \choose k - r}} \cdot \frac{1}{(k -r + 1)!} \sum\limits_{s \in \mathcal{S}_{k - r}^{(a)}} \sum\limits_{\sigma \in \mathcal{P}(s \cup \{ a\})} \prod\limits_{j <_{\sigma} a} (1 - \mathrm{PA}_j(\mathbf{x}_j, t) ) 
    \label{eq:gidef}
\end{equation}

Note that the factor outside of $\mathrm{PA}_a(\mathbf{x}_a,t)$ in equation~\eqref{eq:gidef} depends on the rank $r$ but does not contain $\mathbf{x}_a$ or $\mathrm{PA}_a(\mathbf{x}_a,t)$. So since we are looking at $f_a(\mathbf{x}_a, \mathbf{x}_{-a})$ and $g_{r,a,t}(\mathbf{x}_a, \mathbf{x}_{-a})$ as a function of $\mathbf{x}_a$, we consider this factor a constant in $\mathbf{x}_a$. For an even more simplifying notation: 

\begin{equation}
    C_{r,a,t} := \frac{1}{{k - 1 \choose k - r}} \cdot \frac{1}{(k -r + 1)!} \sum\limits_{s \in \mathcal{S}_{k - r}^{(a)}} \sum\limits_{\sigma \in \mathcal{P}(s \cup \{ a\})} \prod\limits_{j <_{\sigma} a} (1 - \mathrm{PA}_j(\mathbf{x}_j, t) ) , 
    \label{eq:crit}
\end{equation}
thus giving

\begin{equation}
    g_{r,a,t}(\mathbf{x}_a, \mathbf{x}_{-a}) =  \mathrm{PA}_a(\mathbf{x}_a,t) \cdot C_{r,a,t}
\end{equation}

Coming back to the utility function, we get 

\begin{equation}
    f_a(\mathbf{x}_a, \mathbf{x}_{-a}) = \sum\limits_{t = 1}^n \sum\limits_{r = 1}^{k} \alpha_r^2 \cdot  g_{r,a,t}(\mathbf{x}_a, \mathbf{x}_{-a}) \cdot \prod\limits_{\overline{r} < r} \left( 1 - g_{\overline{r},a,t}(\mathbf{x}_a, \mathbf{x}_{-a}) \right)
\end{equation}

Denote by $f_{a,t}$ the summand for each $t \in [1,n]$: 

\begin{equation}
    \begin{gathered}
        f_{a,t}(\mathbf{x}_a, \mathbf{x}_{-a}) = \sum\limits_{r = 1}^{k} \alpha_r^2 \cdot g_{r,a,t}(\mathbf{x}_a, \mathbf{x}_{-a}) \cdot \prod\limits_{\overline{r} < r} \left( 1 - g_{\overline{r},a,t}(\mathbf{x}_a, \mathbf{x}_{-a}) \right),
    \end{gathered}
    \label{eq:fitdef}
\end{equation}
thus giving 

\begin{equation}
    f_a(\mathbf{x}_a, \mathbf{x}_{-a}) = \sum\limits_{t = 1}^n f_{a,t}(\mathbf{x}_a, \mathbf{x}_{-a})
    \label{eq:decomposepersubject}
\end{equation}

First of all, note that an allocation in which an admin does not bid all their budget is not an equilibrium. This follows from the fact that all terms in equation~\eqref{eq:gidef} are non-negative, and $\mathrm{PA}_a(\mathbf{x}_a,t)$ is an increasing function in $\mathbf{x}_a(t)$.~\looseness=-1

\paragraph{A necessary property: } Before proceeding with our proof, we notice that when $\alpha_r = 1, \forall r \in [k]$, we can simplify the expression of the functions $f_{a,t}$ to: 

\begin{equation}
    f_{a,t}(\mathbf{x}_a, \mathbf{x}_{-a}) = 1 - \prod\limits_{r = 1}^k (1 - g_{r,a,t}(\mathbf{x}_a,\mathbf{x}_{-a}))
    \label{eq:simplifyfit}
\end{equation}

Of course, this simplification does not hold if all the terms $\alpha_r$ are not equal to each other. However, we will argue that it is sufficient to show Properties $1$ and $2$ when $\alpha_r = 1, \forall r \in [k]$. To see this, we use the following property: 

\begin{proposition}
    For any $l,v \in \mathbb{N}$ and any functions $h_1, h_2, \cdots, h_l : \mathbb{N}^v \rightarrow \mathbb{R}_{+}$, if for some $x, x' \in \mathbb{N}$ the following conditions hold: 

    \begin{equation}
    \begin{gathered}
        h_1(x) \geq h_1(x') \mbox{ and } \\   
        \sum\limits_{j = 1}^l h_j(x) \geq \sum\limits_{j = 1}^l h_j(x'), 
    \end{gathered}
    \end{equation}
    then for $\omega_1 = 1$ and for any $\omega_2, \cdots, \omega_l \in [0,1]$ the following holds: 

    \begin{equation}
        \sum\limits_{j = 1}^l \omega_j \cdot h_j(x) \geq \sum\limits_{j = 1}^l \omega_j \cdot h_j(x')
    \end{equation}
    \label{prop:trickylittletrick}
\end{proposition}

\begin{corollary}
    We note that the same result holds as in Proposition~\ref{prop:trickylittletrick} if there are two different functions in the first condition: 
    for any $l,v \in \mathbb{N}$ and any functions $h_1, h_2, \cdots, h_l, h_1' : \mathbb{N}^v \rightarrow \mathbb{R}_{+}$, if for some $x, x' \in \mathbb{N}$ the following conditions hold:~\looseness=-1

    \begin{equation}
    \begin{gathered}
        h_1(x) \geq h_1'(x') \mbox{ and } \\   
        h_1(x) + \sum\limits_{j = 2}^l h_j(x) \geq h_1'(x') + \sum\limits_{j = 2}^l h_j(x'), 
    \end{gathered}
    \end{equation}
    then for any $\omega_2, \cdots, \omega_l \in [0,1]$ the following holds: 

    \begin{equation}
        h_1(x) + \sum\limits_{j = 2}^l \omega_j \cdot h_j(x) \geq h_1'(x') + \sum\limits_{j = 2}^l \omega_j \cdot h_j(x')
    \end{equation}
    \label{cor:trickylittletrick}
\end{corollary}

\begin{remark}
    As a remark, Proposition~\ref{prop:trickylittletrick} helps us show that the utility function $f_a$ is increasing in the bid of an admin $a$ for a subject $t$, $\mathbf{x}_a(t)$. This follows from the following properties: 

    \begin{itemize}
        \item First, the function $g_{r,a,t}(\mathbf{x}_a,\mathbf{x}_{-a})$ is increasing in $\mathbf{x}_a(t)$ since $\mathrm{PA}_a(\mathbf{x},t) = 1 - (1-p)^{\mathbf{x}_a(t)}$ is clearly increasing in $\mathbf{x}_a(t)$. This implies that the function $f_{a,t}(\mathbf{x}_a,\mathbf{x}_{-a})$ is also increasing in $\mathbf{x}_a(t)$ from equation~\eqref{eq:simplifyfit} in the case when all $\alpha_r$'s are equal to $1$. 
        \item In the case when all $\alpha_r$'s are not equal to $1$, we employ Proposition~\ref{prop:trickylittletrick} by taking $h_j(\mathbf{x}) = g_{j,a,t}(\mathbf{x}_a,\mathbf{x}_{-a}) \cdot \prod\limits_{\overline{j} < j} \left(1 - g_{\overline{j},a,t}(\mathbf{x}_a,\mathbf{x}_{-a}) \right)$ and $\omega_j = \alpha_j^2$. Then, 

        \begin{equation}
            \sum\limits_{r = 1}^{k} \omega_j \cdot h_j(\mathbf{x}) = \sum\limits_{r = 1}^{k} \alpha_r^2 \cdot g_{r,a,t}(\mathbf{x}_a, \mathbf{x}_{-a}) \cdot \prod\limits_{\overline{r} < r} \left( 1 - g_{\overline{r},a,t}(\mathbf{x}_a, \mathbf{x}_{-a}) \right) = f_{a,t}(\mathbf{x}_a,\mathbf{x}_{-a})
        \end{equation}
    \end{itemize}
\end{remark}

\begin{proof}[Proof of Proposition~\ref{prop:trickylittletrick}]
    We prove Proposition~\ref{prop:trickylittletrick} by induction on $l$. For $l = 1$, the result is trivial given that $h_1(x) \geq h_1(x')$. For $l = 2$, the conditions are that for some $x,x' \in \mathbb{N}^v$, $h_1(x) \geq h_1(x')$ and $h_1(x) + h_2(x) \geq h_1(x') + h_2(x')$. We wish to show that for any $\omega_2 \in [0,1]$, we also get that $h_1(x) + \omega_2 \cdot h_2(x) \geq h_1(x') + \omega_2 \cdot h_2(x')$. If $h_2(x) \geq h_2(x')$, this clearly holds as $\omega_2 \geq 0$. If $h_2(x) < h_2(x')$, we get 

    \begin{equation}
        \begin{gathered}
            h_1(x) + \omega_2 \cdot h_2(x) \geq h_1(x') + \omega_2 \cdot h_2(x') \Leftrightarrow \\ 
            h_1(x) - h_1(x') + \omega_2 \cdot (h_2(x) - h_2(x')) \geq 0 \Leftrightarrow \\ 
            h_1(x) - h_1(x') + \omega_2 \cdot (h_2(x) - h_2(x')) \geq h_1(x) - h_1(x') + h_2(x) - h_2(x')  \geq 0
        \end{gathered}
    \end{equation}
where the last inequality is true from the conditions of the Proposition. Thus, the base case is proved. For the induction case, assume that the statement of the Proposition is true for $l \geq 2$ and we will show that it is also true for $l + 1$. Given the induction hypothesis, we know that for some $x,x' \in \mathbb{N}^v$, $h_1(x) \geq h_1(x')$ and $\sum\limits_{j = 1}^l h_j(x) \geq \sum\limits_{j = 1}^l h_j(x')$. If for all $j \in \overline{2, l + 1}$, $h_j(x) < h_j(x')$, then we get 

\begin{equation}
    \sum\limits_{j = 1}^{l + 1} \omega_j \cdot (h_j(x) - h_j(x')) \geq \sum\limits_{j = 1}^{l + 1} h_j(x) - h_j(x') \geq 0
\end{equation}
since $\omega_j \geq 0, \forall j$ and the unweighted sum is non-negative from the conditions of the Proposition. If, however, $\exists j \in \overline{2,l+1}$ such that $h_j(x) \geq h_j(x')$, then we also know that $h_1(x) + \omega_j \cdot h_j(x) \geq h_1(x') + \omega_j \cdot h_j(x')$. Then, denote by $h_1'(y) = h_1(y) + \omega_j \cdot h_j(y), \forall y \in \mathbb{N}^v$. We know that 

\begin{equation}
    \begin{gathered}
    h_1'(x) \geq h_1'(x') \mbox{ and that } \\
    h_1'(x) + \sum\limits_{\substack{j' \in \overline{2,l+1}, \\j' \neq j}} \omega_{j'} \cdot h_{j'}(x) \geq h_1'(x') + \sum\limits_{\substack{j' \in \overline{2,l+1}, \\ j' \neq j}} \omega_{j'} \cdot h_{j'}(x'),
    \end{gathered}
\end{equation}
    and thus we can apply the induction hypothesis, concluding our proof. The corollary follows immediately by the same proof.~\looseness=-1
\end{proof}

The proof of Lemma~\ref{lemma:nash_kgen-c0} follows the following structure: we use Proposition~\ref{prop:trickylittletrick} to show that if an allocation $\mathbf{x}_a$ has better utility than an allocation $\mathbf{x}_a'$ for a subject slot $t$ when $\alpha_r = 1, \forall r \in [k]$ (meaning that $f_{a,t}(\mathbf{x}_a, \mathbf{x}_{-a}) > f_{a,t}(\mathbf{x}_a', \mathbf{x}_{-a})$), then $\mathbf{x}_a$ has better utility than $\mathbf{x}_a$ for any $\alpha_r \in [0,1]$. This follows by taking $l = k$, $\omega_r = \alpha_r^2, \forall r \in [k]$, and $h_r(x) = g_{r,a,t}(\mathbf{x}_a, \mathbf{x}_{-a}) \cdot \prod\limits_{\overline{r} < r} (1 - g_{\overline{r},a,t}(\mathbf{x}_a, \mathbf{x}_{-a}))$.
Lastly, we will verify the first condition of the Proposition~\ref{prop:trickylittletrick}, that $h_1(x) \geq h_1'(x')$ for each of the Properties $1$ and $2$, below. If that holds, then we conclude that if an allocation satisfies Properties $1$ and $2$ when $\alpha_r = 1, \forall r\in [k]$, then it satisfies Properties $1$ and $2$ for any $\alpha_r \in [0,1], \forall r \in \overline{2,k}$ (knowing that we always set $\alpha_1 = 1$). Finally, if an allocation $\mathbf{x}$ satisfies Properties $1$ and $2$, then it is a Nash equilibrium. 

\paragraph{Proof of Property 1: } Then, we proceed to show Property $1$. Without loss of generality, assume $a = 1$. If there is no subject on which admin $1$ has allocated $0$ budget or no subject on which admin $1$ has allocated $1$ budget, there is nothing to show. Therefore, take two subjects, $t_0$ and $t_1$, such that $\mathbf{x}_1(t_0) = 0$ and $\mathbf{x}_1(t_1) = 1$. We would like to show that admin $1$ cannot increase their utility by moving the unit of budget from $t_1$ to $t_0$. If the allocations of all other admins on subjects $t_0$ and $t_1$ are exactly the same, then admin $i$ has the same utility from allocating $1$ unit of budget on $t_0$ or $t_1$. Moreover, given the assumption that the activation probability is the same across admins, if the allocations of all other admins on subjects $t_0$ and $t_1$ have the same number of $0$s and $1$s, then, again, admin $i$ has the same utility from allocating $1$ unit of budget on $t_0$ or $t_1$. If, however, the allocations of all other admins on subjects $t_0$ and $t_1$ have a different number of $0$s and $1$s, then we know that the number of $0$s in the allocations of $t_1$ is higher than in $t_0$ (excluding admin $1$). This is because of the property stated in the allocation described in Lemma~\ref{lemma:nash_kgen-c0}: the allocation $\mathbf{x}$ minimizes the number of $0$s in each $\mathbf{x}_{\cdot,t}$ (as admin $1$ prioritizes subjects with least bids); therefore, the number of $0$s in the allocations of $t_1$ is higher than in $t_0$ (excluding admin $1$). We are almost done at this point, as we only need the following remark: 
\begin{remark}
    Take the following two allocations of all admins over subject $t$: 

\begin{equation}
    \mathbf{x}_{\cdot,t} = \begin{pmatrix}
\mathbf{x}_{1t}  \\
\mathbf{x}_{2t} \\ 
\vdots \\ 
\mathbf{x}_{kt}
\end{pmatrix} \mbox{ and } \mathbf{x}_{\cdot,t}' = \begin{pmatrix}
\mathbf{x}_{1t}' \\
\mathbf{x}_{2t}' \\ 
\vdots \\ 
\mathbf{x}_{kt}'
\end{pmatrix},
\end{equation}
where $\mathbf{x}_{\cdot,t}$ has the first $j$ coordinates as $1$s and the rest zeros, and $\mathbf{x}_{\cdot,t}'$ has the first $j + 1$ coordinates as $1$s and the rest zeros, for some $j \geq 0$. Then, 

\begin{equation}
    f_{a,t} \left( \mathbf{x}_{\cdot,t} \right) \geq f_{a,t} \left( \mathbf{x}_{\cdot,t}' \right)
\end{equation}
\label{obs:bettertohavemore0s}
\end{remark}

From the assumption that $p_a = p \in [0,1]$ is the same across admins and subjects, we get that Remark~\ref{obs:bettertohavemore0s} is true even for different subjects, and moreover, even for different permutations of the coordinates of $\mathbf{x}_{\cdot,t}$ and $\mathbf{x}_{\cdot,t}'$. (Basically, any allocation that contains $j$ ones and $k - j$ zeros will have better utility for admin $a$ than any allocation that contains $j + 1$ ones and $k - j - 1$ zeros.)

Using Remark~\ref{obs:bettertohavemore0s} iteratively, this shows that the utility of admin $1$ cannot increase if it moves a unit of budget from a subject with fewer allocations of budget units from the other admins to a subject with more allocations of budget units from other admins. 

\begin{proof}[Proof of Remark~\ref{obs:bettertohavemore0s}]

Finally, to prove Remark~\ref{obs:bettertohavemore0s}, we simply use the fact that, if we take two allocations of admin $j$ as $\mathbf{x}_j$ and $\mathbf{x}_j'$ that only differ on coordinate $t$ (for some $t \in [n]$), for example $\mathbf{x}_j(t) = 0$ and $\mathbf{x}_j'(t) = 1$, then $\mathrm{PA}_j(\mathbf{x}_j, t) \leq \mathrm{PA}_j(\mathbf{x}_j', t)$. This is true since $\mathrm{PA}_j(\mathbf{x}_j, t) = 0$ and $\mathrm{PA}_j(\mathbf{x}_j', t) = p_j \geq 0$. Thus, we get that 

\begin{equation}
        \begin{gathered}
            \mathrm{PA}_j(\mathbf{x}_j, t) \leq \mathrm{PA}_j(\mathbf{x}_j', t) \Rightarrow  \\ 
            1 - \mathrm{PA}_j(\mathbf{x}_j, t) \geq 1 - \mathrm{PA}_j(\mathbf{x}_j', t) \Rightarrow \\ 
            C_{r,a,t}(\mathbf{x}_{-a}) \geq C_{r,a,t}(\mathbf{x}_{-a}'), \forall r\in [k] \Leftrightarrow \\ 
            g_{r,a,t}(\mathbf{x}_a,\mathbf{x}_{-i}) \geq g_{r,a,t}(\mathbf{x}_a,\mathbf{x}_{-a}'), \forall r \in [k], \mbox{ where } \mathbf{x}_j \in \mathbf{x}_{-a} \mbox{ and } \mathbf{x}_j' \in \mathbf{x}_{-a}' \Rightarrow \\ 
            1 - g_{r,a,t}(\mathbf{x}_a,\mathbf{x}_{-a}) \leq 1 - g_{r,a,t}(\mathbf{x}_a,\mathbf{x}_{-a}'), \forall r \in [k] \Rightarrow \\ 
            \prod\limits_{r = 1}^k (1 - g_{r,a,t}(\mathbf{x}_a,\mathbf{x}_{-a})) \leq \prod\limits_{r = 1}^k (1 - g_{r,a,t}(\mathbf{x}_a,\mathbf{x}_{-a}')) \Rightarrow \\ 
            1 - \prod\limits_{r = 1}^k (1 - g_{r,a,t}(\mathbf{x}_a,\mathbf{x}_{-a})) \geq 1 - \prod\limits_{r = 1}^k (1 - g_{r,a,t}(\mathbf{x}_a,\mathbf{x}_{-a}')) \Rightarrow \\ 
            f_{a,t}(\mathbf{x}_a,\mathbf{x}_{-a}) \geq 
            f_{a,t}(\mathbf{x}_a,\mathbf{x}_{-a}'),
        \end{gathered}
\end{equation}
From the fourth inequality above, we also deduce that the first condition for the Proposition~\ref{prop:trickylittletrick} is satisfied by taking $h_1 = g_{1,a,t}$. Noting that $\mathbf{x}_j$ and $\mathbf{x}_j'$ differ on just one coordinate and we showed that the allocation that contained one more budget unit on subject $t$ has a better utility for admin $i$, this concludes the proof of the remark. As mentioned above, using Proposition~\ref{prop:trickylittletrick}, the remark holds for any $\alpha_r \in [0,1], \forall r \in \overline{2,k}$ as well. 
\end{proof}

\paragraph{Proof of Property 2: } Next, we that an admin does not prefer to aggregate his spent budget across any subject slots. Again, without loss of generality, take admin $a = 1$ and take two allocations over two subject slots, from an allocation $\mathbf{x}$ described in Lemma~\ref{lemma:nash_kgen-c0} (and we will argue for any number at the end, as it easily generalizes):~\looseness=-1

\begin{equation}
    \mathbf{x}_{\cdot,t} = \begin{pmatrix}
        \mathbf{x}_{1t} \\
        \vdots \\
        \mathbf{x}_{kt}
    \end{pmatrix} \mbox{ and } \mathbf{x}_{\cdot,t'}' = \begin{pmatrix}
        \mathbf{x}_{1t'}' \\
        \vdots \\
        \mathbf{x}_{kt'}'
    \end{pmatrix}
\end{equation}

An admin can aggregate his budget spent on $t$ and $t'$ on either of them, or a new subject $t''$. For our proof, it does not matter, since the property we use is that the number of $0$s among the others admins' bids for any of the subjects $t,t',t''$ does not differ by more than one. Thus, let's assume that the admin aggregates his budget from $t$ and $t'$ onto subject $t$ (instead of spending $\mathbf{x}_{1t}$ and $\mathbf{x}_{1t'}$, he spends $\mathbf{x}_{1t} + \mathbf{x}_{1t'}$ and $0$ on subjects $t$ and $t'$, respectively). 
If either $\mathbf{x}_{1t}$ or $\mathbf{x}_{1t'}'$ are equal to $0$, then there is nothing to aggregate since we reduce to Property $1$, so we only deal with the case $\mathbf{x}_{1t} = \mathbf{x}_{1t'}' = 1$. Since the distribution of $0$s and $1$s are such that the number of $0$s on each subject slot of $\mathbf{x}$ is minimized, a pair of allocations of two subject slots cannot differ by more than $1$ in their number of $0$s. In addition, given the assumption that the relevance probability is the same across subject slots and admins, the order of the ones and zeros does in the allocations does not matter.~\looseness=-1

Case 1: $\mathbf{x}_{\cdot, t}$ and  $\mathbf{x}_{\cdot, t'}'$ have the same number of zeros and ones. Given that the order does not matter, $f_{1,t} \left(\mathbf{x}_{\cdot, t} \right) = f_{1,t'} \left(\mathbf{x}_{\cdot, t'} \right)$. Aggregating the spent budget into $\mathbf{x}_{\cdot,t}$ would result into an allocation like

\begin{equation}
    \mathbf{x}_{\cdot,t}^{\textrm{a}} = \begin{pmatrix}
        \mathbf{x}_{1t} + \mathbf{x}_{1t'}' \\
        \mathbf{x}_{2t}
        \vdots \\
        \mathbf{x}_{kt}
    \end{pmatrix}  = \begin{pmatrix}
        2 \\
        \mathbf{x}_{2t}
        \vdots \\
        \mathbf{x}_{kt}
    \end{pmatrix} 
    \label{eq:xagg}
\end{equation}
as $\mathbf{x}_{1t} = \mathbf{x}_{1t'}' = 1$.

Of course, the admin can aggregate their spent budget into either allocation $\mathbf{x}_{\cdot,t}$ or $\mathbf{x}_{\cdot,t'}$, but that does not change the utility contribution in this case. We are thus left to show that 

\begin{equation}
    2 \cdot f_{1,t} \left(\mathbf{x}_{\cdot,t} \right) \geq f_{1,t} \left(\mathbf{x}_{\cdot,t}^{\textrm{a}} \right),
\end{equation}
which would also imply that the first condition of Corollary~\ref{cor:trickylittletrick} is satisfied, by taking $h_1 = 2 f_{1,t}$ and $h_1' = f_{1,t}$.

Since all the coordinates except the first one are the same in $\mathbf{x}_{\cdot, t}$ and $\mathbf{x}_{\cdot, t}^{\textrm{a}}$, all the $C_{r,a,t}$ terms  in the utility function expression will be same. We also show that 

\begin{equation}
\begin{gathered}    
    2 \cdot \mathrm{PA}_1(\mathbf{x}_{\cdot,t}, t) \geq \mathrm{PA}_1(\mathbf{x}_{\cdot,t}^{\textrm{a}}, t) \Leftrightarrow \\ 
    2 \cdot p \geq 1 - (1 - p)^2 \Leftrightarrow \\ 
    2 \geq 1 + (1 - p),
\end{gathered}
\label{eq:2pa1ineq}
\end{equation}
which is true since $p \in [0,1]$. This also implies that $g_{1,a,t}(\mathbf{x}_{\cdot,t}) \geq g_{1,a,t}(\mathbf{x}_{\cdot,t}^a)$.
Finally, 

\begin{equation}
    \begin{gathered}
        1 - \mathrm{PA}_1(\mathbf{x}_{\cdot,t}, t) \cdot C_{r,1,t}  \geq 1 - \mathrm{PA}_1(\mathbf{x}_{\cdot,t}^{\textrm{a}}, t) \cdot C_{r,1,t} \Leftrightarrow \\
        \mathrm{PA}_1(\mathbf{x}_{\cdot,t}, t) \cdot C_{r,1,t} \leq \mathrm{PA}_1(\mathbf{x}_{\cdot,t}^{\textrm{a}}, t) \cdot C_{r,1,t} \Leftrightarrow \\
        \mathrm{PA}_1(\mathbf{x}_{\cdot,t}, t)  \leq \mathrm{PA}_1(\mathbf{x}_{\cdot,t}^{\textrm{a}}, t) \Leftrightarrow \\
        1 - (1 - p) \leq 1 - (1 - p)^2 \Leftrightarrow \\ 
        (1 -p)^2 \leq (1 - p),
    \end{gathered}
    \label{eq:critpaineq}
\end{equation}
which is true for any $r \in [k]$ since $p \in [0,1]$.
We note that these last two inequalities remains true if we replace $2$ by $N$, for any $N \geq 2$. Combining inequalities~\eqref{eq:2pa1ineq} and~\eqref{eq:critpaineq} in the formula for $f_{1,t}$ and noting that all terms are non-negative, we obtain the desired result with Corollary~\ref{cor:trickylittletrick}.

Case 2: $\mathbf{x}_{\cdot, t}$ and  $\mathbf{x}_{\cdot, t}'$ differ by $1$ in the number of ones they contain. It is sufficient to consider the case where $\mathbf{x}_{\cdot, t}$ has $j$ ones and  $\mathbf{x}_{\cdot, t}'$ has $j + 1$ ones, for some $j \geq 1$. Given Remark~\ref{obs:bettertohavemore0s}, it is sufficient to show

\begin{equation}
    2 \cdot f_{1,t'} \left( \mathbf{x}_{\cdot,t'}' \right) \geq f_{1,t} \left( \mathbf{x}_{\cdot,t}^{\textrm{a}} \right) 
\end{equation}
with $\mathbf{x}_{\cdot,t}^{\textrm{a}}$ defined as in equation~\eqref{eq:xagg} (since $\cdot f_{1,t} \left( \mathbf{x}_{\cdot,t} \right) \geq\cdot f_{1,t'} \left( \mathbf{x}_{\cdot,t'}' \right)$). We will show that for each $r \in [k]$:

\begin{equation}
\begin{gathered}
    2 \cdot \mathrm{PA}_1(\mathbf{x}_{\cdot,t'}', t') \cdot C_{r,1,t}' \geq \mathrm{PA}_1(\mathbf{x}_{\cdot,t}^{\textrm{a}}, t) \cdot C_{r,1,t},
\end{gathered}
\end{equation}
where the terms $C_{r,a,t}$ are defined as in equation~\eqref{eq:crit} and that

\begin{equation}
\begin{gathered}
    2 \cdot \mathrm{PA}_1(\mathbf{x}_{\cdot,t'}', t') \cdot C_{r,1,t}'   \cdot \prod\limits_{\overline{r} < r} \left( 1 - \mathrm{PA}_1(\mathbf{x}_{a,t'}', t') \cdot C_{\overline{r},1,t}'  \right) \geq \\
    \mathrm{PA}_1(\mathbf{x}_{\cdot,t}^{\textrm{a}}, t) \cdot C_{r,1,t}   \cdot \prod\limits_{\overline{r} < r} \left( 1 - \mathrm{PA}_1(\mathbf{x}_{\cdot,t}^{\textrm{a}}, t) \cdot C_{\overline{r},1,t}  \right)
\end{gathered}
\end{equation}

First of all, we note that the difference between $C_{r,1,t}$ and $C_{r,1,t}'$ is that there is an extra $1$ in the allocation of $\mathbf{x}_{\cdot,t}'$ than in $\mathbf{x}_{\cdot,t}$. This extra $1$ contributes to the change between $C_{r,1,t}$ and $C_{r,1,t}'$ in the following way: let's say that $C_{r,1,t}$ is composed of two types of terms, the terms in which the extra one shows up in a factor $(1 - \mathrm{PA}_j(\cdot))$, and the terms in which it does not. Thus, let's write $C_{r,1,t}$ as $C_{r,1,t} = C_{r,1,t}^{(j)} + C_{r,1,t}^{(-j)}$. We now note that $C_{r,1,t}' = C_{r,1,t}^{(j)} \cdot (1 - p) + C_{r,1,t}^{(-j)}$, since the terms that did not contain the extra $1$ coordinate did not change, while those that did simply changed from $1$ to $1 - p$. This also shows that $C_{r,1,t}' \leq C_{r,1,t}$ for $p \in [0,1]$. 

Thus, for any $\overline{r} \in [k]$, 

\begin{equation}
\begin{gathered}
    1 - \mathrm{PA}_1(\mathbf{x}_{\cdot,t'}',t') \cdot C_{\overline{r},1,t}'  \geq 1 - \mathrm{PA}_1(\mathbf{x}_{\cdot,t}^{(a)},t) \cdot C_{\overline{r},1,t}  \Leftrightarrow \\ 
    \mathrm{PA}_1(\mathbf{x}_{\cdot,t'}',t') \cdot C_{\overline{r},1,t}' \leq \mathrm{PA}_1(\mathbf{x}_{\cdot,t}^{(a)},t) \cdot C_{\overline{r},1,t} \Leftrightarrow \\
    \mathrm{PA}_1(\mathbf{x}_{\cdot,t'}',t')  \leq \mathrm{PA}_1(\mathbf{x}_{\cdot,t}^{(a)},t) \Leftrightarrow \\
    1 - (1 - p) \leq 1 - (1 - p)^2 \Leftrightarrow \\ 
    1 - p \leq (1 - p)^2,
    \end{gathered}
\end{equation}
which is true since $p \in [0,1]$. Finally, we also show that 

\begin{equation}
\begin{gathered}
     2 \cdot \mathrm{PA}_1(\mathbf{x}_{\cdot,t'}',t') \cdot C_{r,1,t}'   \geq  \mathrm{PA}_a(\mathbf{x}_{\cdot,t}^{(a)},t) \cdot C_{r,1,t}   \Leftrightarrow \\ 
     2 (1 - (1 -p) ) \cdot \left( (1 -p) \cdot C_{r,1,t}^{(j)} + C_{r,1,t}^{(-j)} \right) \geq (1 - (1 -p)^2 ) \cdot \left( C_{r,1,t}^{(j)} + C_{r,1,t}^{(-j)}  \right) \Leftrightarrow \\ 
     C_{r,1,t}^{(-j)} \geq C_{r,1,t}^{(j)},
\end{gathered}
\label{eq:2pacritineq}
\end{equation}
which is true for all $r\in [k]$ since the sum of the terms in permutations where $1$ shows up \textit{before} $j$ ($j$ being the coordinate of the extra budget unit in the allocation $\mathbf{x}_{\cdot,t'}'$) is at least equal to the sum of the terms in permutations where $1$ shows up \textit{after} $j$ (each permutation $\sigma$ in which $1$ shows up before $j$ can be `paired' with a permutation $\sigma$ in which $1$ is after $j$ by just switching their position. Thus, by the definition of the utility function, the product contribution coming from $\sigma$ is at most the product contribution coming from $\sigma'$). We note that inequality~\eqref{eq:2pacritineq} holds when we replace $2$ by any $N \geq 2$ since the function $\frac{N \cdot p}{1 - (1 - p)^N}$ is increasing in $N$ (while all other terms do not contain $N$). Similarly as in Case $1$, the conditions for Corollary~\ref{cor:trickylittletrick} hold for $h_1 = 2g_{r,a,t}$ and $h_1' = g_{r,a,t}$. 

\end{proof}

\begin{proof}[Proof of Proposition~\ref{prop:onlynashis01_k2}]

    Assume by contradiction that there is a Nash equilibrium allocation $\mathbf{x}$ that does not satisfy the conditions of Lemma~\ref{lemma:nash_kgen-c0}. First, we know admins always spend all their budget in a Nash equilibrium from the proof of Lemma~\ref{lemma:nash_kgen-c0}. Second, from the same proof we also know that if the bids are all $0$ or $1$, then admins also minimize competition (property $1$). Therefore, the only way in which an allocation $x$ does not fulfill the conditions of Lemma~\ref{lemma:nash_kgen-c0} and might \textit{still} be a Nash equilibrium is if some of the bids are not $0$ or $1$. Thus, assume without loss of generality that there exists a subject of subjects $T \subset [n]$ for which $\mathbf{x}_{at} > 1$ $\forall t \in T$, for an admin $a$. Without loss of generality, assume that $a = 2$ (the second admin). We have two cases:~\looseness=-1 
    
    \begin{itemize}
        \item Case $1$: there is a subject $t \in T$ such that $\mathbf{x}_{1t} = 0$.
        \item Case $2$: there is no subject $t \in T$ such that $\mathbf{x}_{1t} = 0$.
    \end{itemize}
    Let's first tackle Case $1$. In this case, we know that there is a subject $t \in T$ such that $\mathbf{x}_{1t} = 0$. Since $\mathbf{x}_{2t} > 1$ for $t \in T$ and $B_2 \leq n$, there exists a subject $t' \notin T$ such that $\mathbf{x}_{2t'} = 0$. Denote $\mathbf{x}_{2t} := x$ and $\mathbf{x}_{1t'} := 0$. We will show that the configuration $ \begin{pmatrix}
        0 & x' \\ 
        x & 0
    \end{pmatrix}$ is not a Nash equilibrium for $x, x' > 1$. If one of $x$ or $x'$ is equal to $1$, then we know this is not Nash because of Property $2$ in the proof of Lemma~\ref{lemma:nash_kgen-c0}. Assume without loss of generality that $x \leq x'$. We will show that 

    \begin{equation}
        \begin{gathered}
            f_1 \begin{pmatrix}
        0 & x' \\ 
        x & 0
    \end{pmatrix} < f_1 \begin{pmatrix}
        1 & x' - 1 \\ 
        x & 0
    \end{pmatrix}
        \end{gathered}
        \label{eq:util_comparison}
    \end{equation}

    We want to show this for any sequence $(\alpha_r)_r$. However, due to Proposition~\ref{prop:trickylittletrick}, we will argue that it is sufficient to show it for $\alpha_r = 1, \forall r \in [k]$. To see why Proposition~\ref{prop:trickylittletrick} applies, we take as functions $h_r$ each term corresponding to $\alpha_r^2$, for all $r \in [k]$. We also compute the terms corresponding to $\alpha_1^2$ in the utility function, denoting $\overline{p}:= 1- p$ for ease of notation.
    Using the combinatorial formula for the utility, we get that the coefficient of $\alpha_1^2$ in $f_1 \begin{pmatrix}
        0 & x' \\ 
        x & 0
    \end{pmatrix} $ is $1 - \overline{p}^{x'}$, and the coefficient of $\alpha_1^2$ in $f_1 \begin{pmatrix}
        1 & x' - 1 \\ 
        x & 0
    \end{pmatrix} $ is $1 - \overline{p}^{x' - 1} + \frac{1}{2}(1 -\overline{p})\cdot (1 + \overline{p}^{x})$. Since $x \leq x'$ and all terms are non-negative, it is sufficient to show that 

    \begin{equation}
    \begin{gathered}
        1 - \overline{p}^{x'} < 1 - \overline{p}^{x' - 1} + \frac{1}{2}(1 -\overline{p})\cdot (1 + \overline{p}^{x'}) \Leftrightarrow \\ 
        \overline{p}^{x' - 1} \cdot (1 - \overline{p}) < \frac{1}{2} (1 - \overline{p}) \cdot (1 + \overline{p}^{x'}) \Leftrightarrow \\ 
        \overline{p}^{x' - 1} < \frac{1}{2} \cdot (1 + \overline{p}^{x'}),
        \end{gathered}
    \end{equation}
    which is true for $x' > 1$. We have used that $\overline{p} \in (0,1)$. Thus, the conditions for Proposition~\ref{prop:trickylittletrick} would be satisfied. We are left to show that inequality~\eqref{eq:util_comparison} happens for $\alpha_r = 1, \forall r \in [k]$. Writing out the utility functions from their combinatorial form, we get

    \begin{equation}
        f_1 \begin{pmatrix}
        0 & x' \\ 
        x & 0
    \end{pmatrix} = (1 - \overline{p}^{x'}) \cdot (1 + \overline{p}^{x'})
    \end{equation}
    and 
    \begin{equation}
        f_1 \begin{pmatrix}
        1 & x' - 1 \\ 
        x & 0
    \end{pmatrix} = (1 - \overline{p}^{x' - 1}) \cdot (1 + \overline{p}^{x' - 1}) + (1 - \overline{p}) \cdot \left( 1 + \frac{1}{2} \overline{p}(1 + \overline{p}^x)\right)
    \end{equation}

    Again, since $x \leq x'$ and all terms are non-negative, it is sufficient to show 

    \begin{equation}
        \begin{gathered}
            (1 - \overline{p}^{x'}) \cdot (1 + \overline{p}^{x'}) < (1 - \overline{p}^{x' - 1}) \cdot (1 + \overline{p}^{x' - 1}) + (1 - \overline{p}) \cdot \left( 1 + \frac{1}{2} \overline{p}(1 + \overline{p}^{x'})\right)
        \end{gathered}
        \label{eq:nastyeq}
    \end{equation}
    Equation~\eqref{eq:nastyeq} simplifies to 

    \begin{equation}
        \begin{gathered}
            \overline{p}^{2x' - 2}(1 + \overline{p}) < 1 +\frac{1}{2} \overline{p} \left( 1 + \overline{p}^{x'}\right)
        \end{gathered}
    \end{equation}
    This is true for $x' > 2$ since $\overline{p}^{2x' - 1} < 1$, $\overline{p}^{2x' - 2} < \overline{p}$, and $\overline{p}^{2x' - 2} < \overline{p}^{x'  + 1}$. For $x' = 2$, the equation simplifies to 

    \begin{equation}
        \overline{p}^2 + \frac{1}{2} \overline{p}^3 < 1 + \frac{1}{2} \overline{p},
    \end{equation}

    We next tackle Case $2$, for which we know that there is no subject $t \in T$ such that $\mathbf{x}_{1t} = 0$. Now, if there exists a subject $t \notin T$ such that $\mathbf{x}_{1t} > 1$, there must also exist a subject $t' \notin T$ such that $\mathbf{x}_{1t'} = 0 $, since $B_1 \leq n$. However, we know that $\mathbf{x}_{2t}, \mathbf{x}_{2t'} \in \{ 0 ,1\}$ as $t,t' \notin T$, by definition of the set $T$. However, none of the configurations $\begin{pmatrix}
        \mathbf{x}_{1t} & 0 \\
        0 & 0 
    \end{pmatrix}, \begin{pmatrix}
        \mathbf{x}_{1t} & 0 \\
        1 & 1 
    \end{pmatrix}, \begin{pmatrix}
        \mathbf{x}_{1t} & 0 \\
        1 & 0 
    \end{pmatrix}, \begin{pmatrix}
        \mathbf{x}_{1t} & 0 \\
        0 & 1 
    \end{pmatrix}$ is an equilibrium for $\mathbf{x}_{1t} > 1$, from Property $2$ of the proof of Lemma~\ref{lemma:nash_kgen-c0} (when the other bids are $0$ or $1$, an admin always wants to split their bid equally). That means that for all subjects $t \notin T$, $\mathbf{x}_{1t} \in \{0,1 \}$. We will also show that for all subjects $t \in T$, $\mathbf{x}_{1t} \in \{0,1 \}$. By contradiction, if there is a subject $t \in T$, $\mathbf{x}_{1t} > 1$, take a subject $t' \notin T$ such that $\mathbf{x}_{2t'} = 0$ (we know it exists since admin's $2$ bids in $T$ are greater than $1$ and $B_2 \leq n$). Then, the configurations $\begin{pmatrix}
        \mathbf{x}_{1t} & \mathbf{x}_{1t'} \\ 
        \mathbf{x}_{2t} & 0
    \end{pmatrix}$ are not equilibria for $\mathbf{x}_{1t}, \mathbf{x}_{2t} > 1$ and $\mathbf{x}_{1t'} \in \{0,1 \}$, easy to show by computing the utility function in its combinatorial form and using $\overline{p} \in (0,1)$. Thus, we have shown that for all subjects $t$, $\mathbf{x}_{1t} \in \{0,1\}$. However, we now have a contradiction, since by Property $2$, $\mathbf{x}_{2t} \in \{0,1\}$ as well, which we assumed otherwise for $t \in T$.~\looseness=-1

\end{proof}
    
\end{document}